\documentclass[sn-basic]{sn-jnl}% APA Reference Style 
\usepackage{graphicx}%
\usepackage{multirow}%
\usepackage{amsmath,amssymb,amsfonts}%
\usepackage{amsthm}%
\usepackage{mathrsfs}%
\usepackage[title]{appendix}%
\usepackage{xcolor}%
\usepackage{textcomp}%
\usepackage{manyfoot}%
\usepackage{booktabs}%
\usepackage{algorithm}%
\usepackage{algorithmicx}%
\usepackage{algpseudocode}%
\usepackage{listings}%
\DeclareMathAlphabet      {\mathbfit}{OML}{cmm}{b}{it}

\theoremstyle{thmstyleone}%
\newtheorem{theorem}{Theorem}%  meant for continuous numbers
\newtheorem{proposition}[theorem]{Proposition}% 

\theoremstyle{thmstyletwo}%

\theoremstyle{thmstylethree}%

\begin{document}

\title[Article Title]{Cluster weighted models for functional data\footnote{Submitted to Machine Learning, 12 December 2023}}


\author*[1]{\fnm{Cristina} \sur{Anton}} \email{popescuc@macewan.ca}
\equalcont{Corresponding Author: Cristina Anton}
\author[1]{\fnm{Iain} \sur{Smith}}\email{smithi23@mymacewan.ca}

\affil*[1]{\orgdiv{Department of Mathematics and Statistics}, \orgname{MacEwan University}, \orgaddress{\street{10700 – 104 Avenue}, \city{Edmonton}, \postcode{T5J 4S2}, \state{AB}, \country{Canada}}}


\abstract{We propose a method, funWeightClust,  based on a family of parsimonious models for clustering heterogeneous functional linear regression data.  These models extend cluster weighted models to functional data, and they allow for multivariate functional responses and predictors. The proposed methodology follows the approach used by the the functional high dimensional data clustering (funHDDC) method. We construct an expectation maximization (EM) algorithm for parameter estimation.  Using simulated and benchmark data we show that funWeightClust outperforms  funHDDC and several two-steps clustering methods. We also use funWeightClust  to analyze traffic patterns in Edmonton, Canada.}

\keywords{Model based clustering, Cluster weighted models, Functional linear regression, EM algorithm, Multivariate functional responses, Multivariate functional principal component analysis }



\maketitle
%%%%%%%%%%%%%%%%%%%%%%%%%%%%
\section{Introduction}\label{sec1} 
%%%%%%%%%%%%%%%%%%%%%%%%%%%%%%%%%%
%%%%%%%%%%%%%%%%%%%%%%%%%%%%%%%%%%%%%
%%%%%%%%%%%%%%%%%%%%%%%%%%%%%%%%
The rapid and extensive development of data-intensive technologies has lead to an increased demand for efficient methods to analyze complex data.   Internet of Things (IoT) embedded devices, such as cars, smart watches, and thermostats, record a lot of functional data  for which the observations are represented by functions \citep{RamsaySilverman:2006}. Examples of functional data are the average  daily temperature over a period of time, the daily stock trading prices in financial markets, and the signal recorded in an electrocardiogram  (ECG).  For complex applications many times it is necessary to start by using unsupervised learning algorithms to identify homogeneous groups of data, i.e. clustering the data. Here we propose a new model based clustering method,  funWeightClust, for data that have a functional linear regression relationship between the variables.
\begin{figure}[h!]
\begin{center}
\includegraphics[width=12cm,height=8cm]{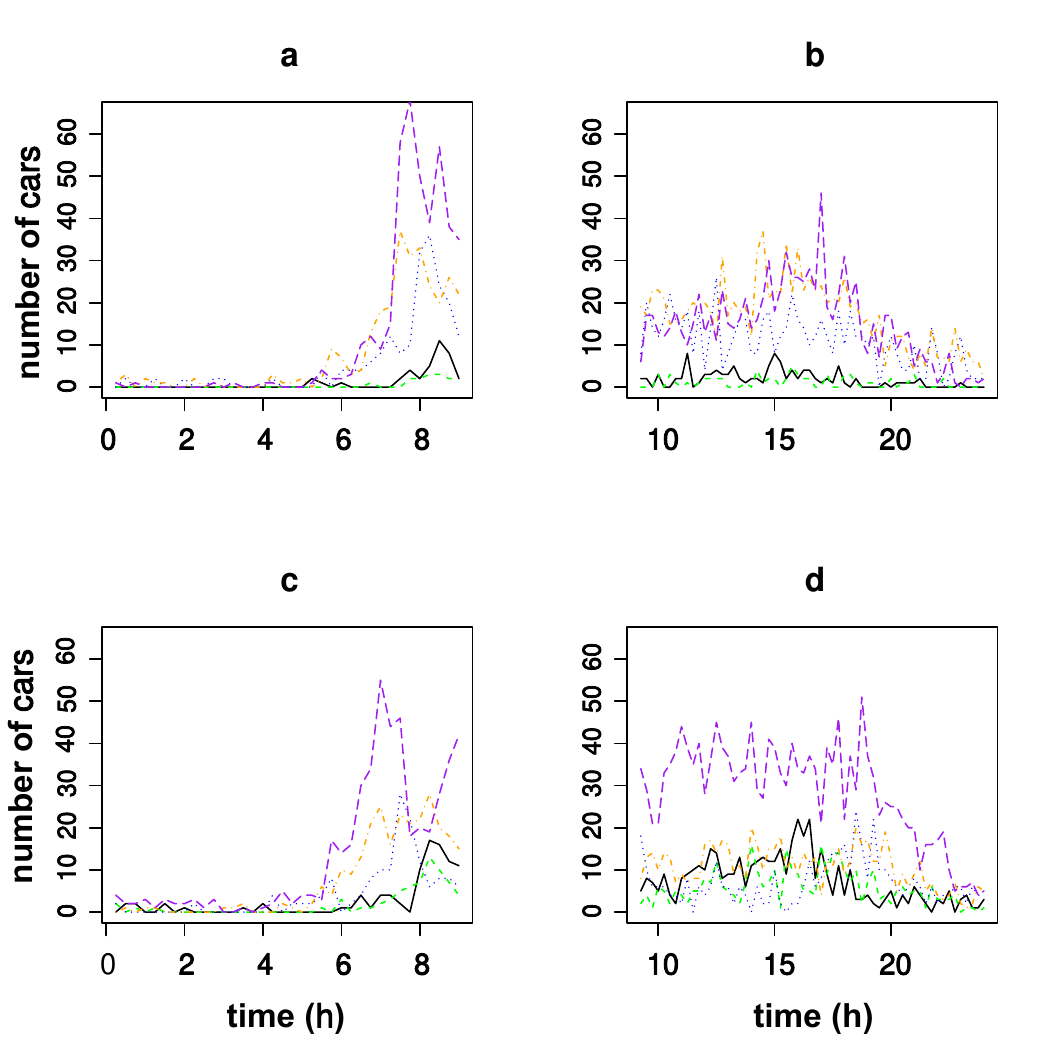}
\end{center}
  \caption{Edmonton traffic data: cars going 5-10 km/h  under the speed limit  a) from 12 a.m. to 8 a.m. b) from 8:15 a.m. to 11:45 p.m.;  cars going 0-5 km/h over the speed limit  c) from 12 a.m. to 8 a.m. d) from 8:15 a.m. to 11:45 p.m.. Roads  with a maximum speed of 30 km/h (black plain lines), 40 km/h  (green dashed lines), 50 km/h (blue dotted lines), 60 km/h (orange dot-dashed lines) and 70 km/h (purple dashed lines).}
\label{fig0}
\end{figure}   

This work was motivated by the study of traffic flow in Edmonton, Canada. The city of Edmonton collects a large amount of data about traffic flow that characterizes road activity. In Figure \ref{fig0} a,b we plot a few curves corresponding to car counts going 5-10 km/h  under the speed limit, and in  Figure \ref{fig0} c, d  we plot curves for car counts going 0-5 km/h over the speed limit. The data is recorded every 15 minutes and we use different colors and line types to distinguish between the curves corresponding to data collected on roads with various maximum speed limits.  The morning traffic rush from 12 a.m. to 8 a.m. is included in Figure \ref{fig0} a, c and the afternoon traffic from 8:15 a.m. to 11:45 p.m. in   Figure \ref{fig0} b, d.  An initial step of traffic analysis is to partition the data in homogeneous groups. The afternoon traffic flow depends on the morning traffic flow, and functional linear regression can model this relationship. Given the nature of the traffic data we have to consider a model that allows for multivariate functional  response and predictors, and, to the best of our knowledge, none of the existing clustering methods (\citealp{YaoFuLee:2011}, \citealp{CondeTavakoliEzer:2021},  \citealp{WangHuangWuYao:2016}, \citealp{Chamroukhi:2016}) deals with multivariate functional responses. 

Time series are also functional data, and special models are used for time series analysis (e.g. the recent work in \citealp{DamaSinoquet:2023}), but here we focus on functional linear regression models.  Our approach extends multivariate cluster weighted models \citep{DangPunzoMcNicholasIngrassiaBrowne:2017} to functional data, based on the framework used by the funHDDC method  (\citealp{ BouveyronJacques:2011}, \citealp{SchmutzJacquesBouveyronChezeMartin:2020}). We assume that the data are collected on an one-dimensional interval and with a regular sampling scheme from the pairs  $(\mathbfit{Y}_1,\mathbfit{X}_1),\ldots, (\mathbfit{Y}_n,\mathbfit{X}_n) $ of multivariate curves. For example, for the Edmonton traffic data  the bi-dimensional covariate or predictors curves $\mathbfit{X}_i$ correspond to the morning rush and are plotted in Figure \ref{fig0} a,c, and the bi-dimensional response curves $\mathbfit{Y}_i$ correspond to the afternoon traffic illustrated in Figure \ref{fig0} b,d. We want to cluster the observations in homogeneous groups and we assume that clusterwise the data come from a functional linear regression model. 

One of the difficulties when clustering functional data is that these data live in an infinite dimensional space and the notion of probability density function generally does not exist  \citep{DelaigleHall:2010}. Thus model-based methods are not directly available for functional data, so we follow the same approach used  for the funHDDC method (\cite{ BouveyronJacques:2011}, \cite{ JacquesPreda:2013}, \cite{SchmutzJacquesBouveyronChezeMartin:2020}). We first  decompose the functional data in a basis of functions (such as Fourier series, B-splines, etc.), then we construct a probabilistic  model for the basis coefficients. 
  
The main contributions of this paper are: 
\begin{itemize}
\item We propose a model based clustering method, funWeightClust,  based on functional linear regression models with multivariate functional  response and predictors.
\item funWeightClust extends the approach used for the funHDDC (\citealp{ BouveyronJacques:2011}, \citealp{SchmutzJacquesBouveyronChezeMartin:2020}) to clustering functional data that include covariates. We use multivariate functional principal component analysis (MFPCA) \citep{JacquesPreda2:2014} and we assume that the scores have  multivariate normal distributions. 
\item We add flexibility to the model by including also the distributions for the covariates, so the proposed approach is also an extension of the cluster weighted models used for multivariate data \citep{DangPunzoMcNicholasIngrassiaBrowne:2017}. 
\item We propose a variant of the Expectation-Maximization (EM) algorithm that allows to learn the parameters of our model. 
\item We consider several parsimonious models. 
\end{itemize}
 
The paper is organized as follows. Related work is reviewed in Section \ref{sectlit}.  The cluster weighted models for functional data are constructed in Section \ref{section2}. In Section \ref{section3} a learning algorithm to estimate the model parameters is derived and computational details are presented.  Next, Section \ref{section4} includes experimental results for both simulated and real-world data and a discussion of the performance of the proposed approach. Finally, in Section \ref{section5} we provide concluding remarks and future directions. 
%%%%%%%%%%%%%%%%%%%%%%%%%%%%%%%
%%%%%%%%%%%%%%%%%%%%%%%%%%%%%%%%%%%%
%%%%%%%%%%%%%%%%%%%%%%%%%%%%%%%%%%

\section{Related work}
\label{sectlit}
Clustering techniques that account for regression relationships between multivariate data can be also applied to functional data if we use one of the following two approaches \citep{JacquesPreda:2014}. The first approach is a raw-data clustering,  and it consists of directly applying multivariate clustering techniques to the  finite discretizations of the functions. The second approach is to use a two-step method  and first do a decomposition of the functional data in a basis of functions (such as Fourier series, B-splines, etc.), and then directly apply multivariate clustering methods to  the basis coefficients. Next, we will summarize some of the model based clustering methods for multivariate data with linear regression relationships between the variables.

Finite mixture of regression (FMR) \citep{DesarboCron:1988} are mixture based methodologies that deal with clusterwise linear regression. They are implemented in the {\it{flexmix}} R package \citep{Leisch:2004} and the implementation allows multivariate uncorrelated responses. However, as in the classical regression analysis, the model assumes that the covariates are fixed and the assignment of the data to clusters is not dependent on the distribution of the covariates.
 
Cluster weighted models (CWM) were first introduced in \cite{Gershenfeld:1997}  and they take into account the distribution of the covariates. In the simplest form it is assumed that both the response and the covariates have normal distributions. Some extensions of this methodology considers  t-distributions \citep{IngrassiaMinottiVittadini:2012}, contaminated normal distributions \citep{PunzoMcNicholas:2017},  and various types of response variables and  covariates of mixed-type (\citealp{IngrassiaPunzoVittadiniMinotti:2015},  \citealp{PunzoIngrassia:2016}). The R package {\it flexCWM} \citep{MazzaPunzoIngrassia:2018} implements models for which the conditioned response variable has some of the most common distributions of the exponential family or the t-distribution. However, these CWMs consider univariate responses, so they cannot be applied to finite discretization of functions or the coefficients corresponding to the decomposition of the functional data in a basis of functions. Very few CWMs consider multivariate responses (\citealp{PunzoMcNicholas:2017}, \citealp{DangPunzoMcNicholasIngrassiaBrowne:2017}).
 
 In \citep{DangPunzoMcNicholasIngrassiaBrowne:2017} a CWM with multivariate correlated responses is constructed. We assume that the response vector $\mathbfit{Y}:\Omega\rightarrow \mathbb{R}^d$, the vector of covariates $\mathbfit{X}:\Omega\rightarrow \mathbb{R}^p$, $\Omega$ is partitioned into $G$ distinct groups, $\Omega=\Omega_1\cup \ldots \cup \Omega_G$, and 
 $$
 \mathbfit{X}|\Omega_g\sim N(\boldsymbol{\mu}_X,\boldsymbol{\Sigma}_X), \quad \mathbfit{Y}|\mathbfit{X}=\mathbfit{x}, \Omega_g\sim N(\mathbfit{B}_g^\top \mathbfit{x}^*, \boldsymbol{\Sigma}_Y), \quad \mathbfit{x}^*=(1,\mathbfit{x}^\top)^\top,
 $$
 where $\mathbfit{B}_g$ is the $(p+1)\times d$ matrix of regression coefficients. The joint probability $p(\mathbfit{x},\mathbfit{y})$ of $\mathbfit{X}$, $\mathbfit{Y}$ can be written as
 \begin{equation}
 p(\mathbfit{x},\mathbfit{y})=\sum_{g=1}^G \pi_g\phi(\mathbfit{y};\mathbfit{B}_g^\top x^*, \boldsymbol{\Sigma}_Y)\phi(\mathbfit{x}; \boldsymbol{\mu}_X,\boldsymbol{\Sigma}_X),
 \label{CWM1}
 \end{equation}
 where $\phi$ represent the density of the Gaussian distribution, and $\pi_g>0$ are the mixing weights with $\sum_{g=1}^G \pi_g=1$.  Here we consider this type of model clusterwise for the coefficients of functional data.
 
Recently finite mixtures of matrix-variate regressions have been considered by \cite{MelnykovZhu:2019} and matrix-variate CWMs are constructed in \cite{TomarchioMcNicholasPunzo:2021}. Moreover, the multivariate response CWM \eqref{CWM1} was extended for the family of models with skewed distributions in \cite{GallaugherTomarchioMcNicholasPunzo:2022}.

In addition to these methods for clustering multivariate data, there is a vast literature in functional linear regression (FLR) (\citealp{RamsaySilverman:2006}, \citealp{FerratyVieu:2006}, \citealp{HorvathKokoszka:2012}). A large variety of FLR models was proposed, but the majority of them consider a scalar response, or a single functional response, and one or more functional predictors. A multivariate FLR model (mFLR) with both multivariate functional response and predictor is constructed in \cite{ChiouYangChen:2016}: 
 \begin{equation}
 Y_k(t)=\sum_{l=1}^p\int_{\mathcal{T}}\beta_{0lk}(s,t)X_l(s)ds+\xi_k(t), \quad k=1,\ldots, d,\label{mFLR}	
 \end{equation}
 where $\{X_l\}_{1\le l\le p}$ and $\{Y_k\}_{1\le l\le d}$ are the predictor and response variables, respectively, $\mathcal{T}$ is a closed interval, $\beta_{0lk}(s,t)$ is the bivariate regression coefficient function, and $\xi_k(t)$ is the random error process. Estimations, predictions, and asymptotic properties of this mFLR model are also discussed in \cite{ChiouYangChen:2016}.   We extend these mFLR models to multivariate functional mixtures models that could be used in applications where observations might come from  an inhomogeneous population with several homogeneous clusters. We assume that for each cluster $c$ we have a model like \eqref{mFLR}, but with cluster specific regression coefficients $\beta_{0lk}^{(c)}(s,t)$.
 
 To the best of our knowledge there are not many papers that consider mixtures of functional linear regression models for clustering. In \cite{Chiou:2012}  conditional of cluster membership, a model similar with \eqref{mFLR}, but with single functional response and predictor ($p=d=1$) and different regression coefficients $\beta^{(c)}(s,t)$ for each cluster $c$, is used for classification and prediction of traffic data. The classification method is based on subspace projection.  
 
 In \cite{YaoFuLee:2011} a functional regression model is used for clustering. In each cluster the model is a simpler version of \eqref{mFLR}, where the response is a scalar $Y(t)=Y\in \mathbb{R}$, there is only one functional predictor ($p=1$) and the univariate regression  coefficient function $\beta^{(c)}(s)$ has specific values for each cluster $c$. The idea of the clustering method is  to use  functional principal components analysis to reduce the functional model to a classical mixture of linear regressions models. 
 
 In \cite{CondeTavakoliEzer:2021}  a cluster-specific model similar with \eqref{mFLR}, with a single functional response and multiple functional predictors ($d=1$, $p\ge 1$), and different regression coefficients $\beta_{0l}^{(c)}(s,t)$, $l=1,\ldots, p$ for each cluster $c$, is used for finding clusters in a gene expression time course data set. The model is fitted using boosting.
 
A mixture constructed with a different type of models, called concurrent  functional linear models, is used in \cite{WangHuangWuYao:2016} to analyze the CO$_2$ emissions - GDP relationship. For each cluster $c$
\begin{equation}
Y(t)=\sum_{l=1}^p\beta_{l}^{(c)}(t)X_l(t)+\xi^{(c)}(t),
\end{equation}
where $\{X_l\}_{1\le l\le p}$ and $Y$ are the predictor and response variables, $\beta_{l}^{(c)}(t)$ is the cluster specific regression coefficient function, and $\xi^{(c)}(t)$ is the random error process. The estimation is done combining the  EM algorithm with kernel regression and  functional principal components analysis. In \cite{Chamroukhi:2016} polynomial,  spline,  and B-spline regression mixtures with a single functional response and multiple functional predictors are used for clustering data. The model is fitted  considering a penalized maximum likelihood and an EM like algorithm.
  %%%%%%%%%%%%%%%%%%%%%%%%%%%%%%
  %%%%%%%%%%%%%%%%%%%%%%%%%%%%%%%%%%
  %%%%%%%%%%%%%%%%%%%%%%%%%%%%%%%%
  %%%%%%%%%%%%%%%%%%%%%%%%%%%%%%%%%%%%%%%%%%
%%%%%%%%%%%%%%%%%%%%%%%%%%%%%%%%%%%%%%%
\section{Multivariate functional cluster weighted  model }
\label{section2}
%%%%%%%%%%%%%%%%%%%%%%%%%%%%
%%%%%%%%%%%%%%%%%%%%%%%%%%%%%%%%%
For any compact interval  $\mathcal{T}$ in $\mathbb{R}$, we consider the Hilbert space $L^2(\mathcal{T})$ $=\{f:\mathcal{T}\rightarrow \mathbb{R}, \int_{\mathcal{T}} f^2(t)dt<\infty\}$ with the inner product $<f,g>=\int_{\mathcal{T}} f(t)g(t)dt$ and the norm $\|f\|=<f,f>^{1/2}$ \citep{RamsaySilverman:2006}.

We assume that the  $n$ $p_Y$-variate response curves $\{{\mathbfit{Y}}_1,\ldots,{\mathbfit{Y}}_n\}$ are independent realizations of a $L^2$- continuous stochastic process $\mathbfit{Y}=\{{\mathbfit{Y}}(t)\}_{t\in \mathcal{T}_Y}$ $=\{(Y^1(t),\ldots,Y^{p_Y}(t))^\top\}_{t\in\mathcal{T}_Y}\in \mathbb{H}_Y$, where $ \mathcal{T}_Y\subset \mathbb{R}$ is a compact interval  and $\mathbb{H}_Y:=\{\mathbfit{f}=(f_1, \ldots,f_{p_Y})^\top:\mathcal{T}_Y\rightarrow \mathbb{R}^{p_Y}, f_i\in L^2(\mathcal{T}_Y), i=1, \ldots, p_Y\}$ is a Hilbert space with the inner product $<\mathbfit{f},\mathbfit{g}>_{\mathbb{H}_Y}=\sum_{l=1}^{p_Y}<f_l,g_l>$ and the norme $\|\mathbfit{f}\|_{\mathbb{H}_Y}=<\mathbfit{f},\mathbfit{f}>_{\mathbb{H}_Y}^{1/2}$.

Similarly we assume that the  $n$ $p_X$-variate covariate curves $\{{\mathbfit{X}}_1,\ldots,{\mathbfit{X}}_n\}$ are independent realizations of a $L^2$- continuous stochastic process $\mathbfit{X}=\{{\mathbfit{X}}(t)\}_{t\in \mathcal{T}_X}$ $=\{(X^1(t),\ldots,X^{p_X}(t))^\top\}_{t\in \mathcal{T}_X}\in \mathbb{H}_X$, where $ \mathcal{T}_X\subset \mathbb{R}$ is a compact interval  and $\mathbb{H}_X:=\{\mathbfit{f}=(f_1, \ldots,f_{p_X})^\top:\mathcal{T}_X\rightarrow \mathbb{R}^{p_X}, f_i\in L^2(\mathcal{T}_X), i=1, \ldots, p_X\}$ is a Hilbert space with the inner product $<\mathbfit{f},\mathbfit{g}>_{\mathbb{H}_X}=\sum_{j=1}^{p_X}<f_j,g_j>$ and the norme $\|\mathbfit{f}\|_{\mathbb{H}_X}=<\mathbfit{f},\mathbfit{f}>_{\mathbb{H}_X}^{1/2}$.

For each pair of curves $({\mathbfit{Y}}_i, {\mathbfit{X}}_i)$ we have access to a finite set of values $y_{i}^{s_Y}(t_{i1}^Y)\ldots,y_{i}^{s_Y}(t_{im_i}^Y)$, $x_{i}^{s_X}(t_{i1}^X)\ldots,x_{i}^{s_X}(t_{in_i}^X)$, where $t_{i1}^Y<t_{i2}^Y<\cdots<t_{im_i}^Y$, $t_{i1}^X<t_{i2}^X<\cdots<t_{in_i}^X$, $t_{ij}^Y\in \mathcal{T}_Y$, $t_{il}^X\in \mathcal{T}_X$, $j=1,\ldots, m_i$, $l=1,\ldots, n_i$, $s_Y=1,\ldots, p_Y$, $s_X=1,\ldots, p_X$, $i=1,\ldots, n$. To reconstruct the functional form of the data we assume that the curves belong to a finite dimensional space, and we have:
\begin{align}
Y_i^l(t)&=\sum_{r=1}^{R_l^Y} c_{Y,ir}^l\xi_{Y,r}^l(t), \quad X_i^j(t)=\sum_{r=1}^{R_j^X} c_{X,ir}^j\xi_{X,r}^j(t).\label{eqfirst}
\end{align} 
Here $\{\xi_{Y,r}^l\}_{1\le r\le R_l^Y}$ is the basis for the $l^{th}$ components of the multivariate curves $\{{\mathbfit{Y}}_1,\ldots,{\mathbfit{Y}}_n\}$, $c_{Y,ir}^l$ are the coefficients, and $R_l^Y$ is the number of basis functions. Similarly for the covariate curves $\{{\mathbfit{X}}_1,\ldots,{\mathbfit{X}}_n\}$,  $\{\xi_{X,r}^j\}_{1\le r\le R_j^X}$ is the basis for the $j^{th}$ components, $c_{X,ir}^j$ are the coefficients, and $R_j^X$ is the number of basis functions.

Gathering the coefficients and the basis functions we rewrite \eqref{eqfirst} as
\begin{align}
{\mathbfit{Y}}(t)&={\mathbfit{C}_Y}\boldsymbol{\xi}_Y^\top (t),\quad {\mathbfit{Y}}(t)=({\mathbfit{Y}}_1(t),\ldots,{\mathbfit{Y}}_n(t))^\top, \label{eqsecond}\\
{\mathbfit{X}}(t)&={\mathbfit{C}_X}\boldsymbol{\xi}_X^\top (t),\quad {\mathbfit{X}}(t)=({\mathbfit{X}}_1(t),\ldots,{\mathbfit{X}}_n(t))^\top, \label{eqsecondbis}
\end{align}
 with
 \begin{align*}
 \mathbfit{C_Y}&=\begin{pmatrix}
 c_{Y,11}^1&\ldots&c_{Y,1R_1^Y}^1&c_{Y,11}^2&\ldots&c_{Y,1R_2^Y}^2&\ldots& c_{Y,11}^{p_Y}&\ldots&c_{Y,1R_p^Y}^{p_Y}\\
 \vdots&\ddots&\vdots&\vdots&\ddots&\vdots&\ldots&\vdots&\ddots&\vdots\\
 c_{Y,n1}^1&\ldots&c_{Y,nR_1^Y}^1&c_{Y,n1}^2&\ldots&c_{Y,nR_2^Y}^2&\ldots &c_{Y,n1}^{p_Y}&\ldots&c_{Y,nR_p^Y}^{p_Y}
 \end{pmatrix},\\
 \boldsymbol{\xi}_Y(t)&=\begin{pmatrix}
 \xi_{Y,1}^1(t)&\ldots&\xi_{Y,R_1^Y}^1&0&\ldots&0&\ldots& 0&\ldots&0\\
 0&\ldots&0&\xi_{Y,1}^2(t)&\ldots&\xi_{Y,R_2^Y}^2(t)&\ldots& 0&\ldots&0\\
 \vdots&\ddots&\vdots&\vdots&\ddots&\vdots&\ldots&\vdots&\ddots&\vdots\\
 0&\ldots&0&0&\ldots&0&\ldots &\xi_{Y,1}^{p_Y}(t)&\ldots&\xi_{Y,R_{p_Y}^Y}^{p_Y}(t)
 \end{pmatrix},\\
 %%%%%%%%%%%%%%%%%%
 \mathbfit{C_X}&=\begin{pmatrix}
 c_{X,11}^1&\ldots&c_{X,1R_1^X}^1&c_{X,11}^2&\ldots&c_{X,1R_2^X}^2&\ldots& c_{X,11}^p&\ldots&c_{X,1R_p^X}^{p_X}\\
 \vdots&\ddots&\vdots&\vdots&\ddots&\vdots&\ldots&\vdots&\ddots&\vdots\\
 c_{X,n1}^1&\ldots&c_{X,nR_1^X}^1&c_{X,n1}^2&\ldots&c_{X,nR_2^X}^2&\ldots &c_{X,n1}^p&\ldots&c_{X,nR_p^X}^{p_X}
 \end{pmatrix},\\
 \boldsymbol{\xi}_X(t)&=\begin{pmatrix}
 \xi_{X,1}^1(t)&\ldots&\xi_{X,R_1^X}^1&0&\ldots&0&\ldots& 0&\ldots&0\\
 0&\ldots&0&\xi_{X,1}^2(t)&\ldots&\xi_{X,R_2^X}^2(t)&\ldots& 0&\ldots&0\\
 \vdots&\ddots&\vdots&\vdots&\ddots&\vdots&\ldots&\vdots&\ddots&\vdots\\
 0&\ldots&0&0&\ldots&0&\ldots &\xi_{X,1}^{p_X}(t)&\ldots&\xi_{X,R_{p_X}^X}^{p_X}(t)
 \end{pmatrix}.
 \end{align*}

 We want to cluster the $n$ observed response and covariate curves $\{({\mathbfit{y}}_1,{\mathbfit{x}}_1),\ldots,({\mathbfit{y}}_n,{\mathbfit{x}}_n)\}$ in $K$ homogeneous groups. We suppose that there exists  a latent variable ${{\mathbfit{Z}}_i}=(Z_{i1},\ldots,Z_{iK})^\top$, associated to each observation $({\mathbfit{y}_i,\mathbfit{x}}_i)$, where $Z_{ik}=1$ if the observation $({\mathbfit{y}_i,\mathbfit{x}}_i)$ belongs to the cluster $k$ and $Z_{ik}=0$ otherwise.  
 We assume that for every $k\in\{1,\ldots,K\}$, given that $Z_{ik}=1$, the observations come from the following model:
 \begin{equation}
 \mathbfit{Y}_i(t)=\boldsymbol{\beta}_{0}^k(t)+\int_{\mathcal{T}_X}\boldsymbol{\beta}^k(t,s)\mathbfit{X}_i(s)ds+\mathbfit{E}^k(t), \quad t\in \mathcal{T}_Y, \quad i=1,\ldots, n. \label{reg1}
 \end{equation}
 Here $\mathbfit{E}^k(t)=(E_{1}^k(t),\ldots, E_{p_Y}^k(t))^T$ is the random error process which is uncorrelated with $\mathbfit{X}_i(s)$ for any $(s,t)\in \mathcal{T}_X \times \mathcal{T}_Y$, and for which we have the expansions
 \begin{equation}
 E_{l}^k(t)=\sum_{r=1}^{R_l^Y}\epsilon_{0,l}^{k,r}\xi_{Y,r}^l(t), \quad l=1,\ldots,p_Y.
 \end{equation}
Let $R^X:=\sum_{j=1}^{p_X} R_j^X$ and $R^Y:=\sum_{l=1}^{p_Y} R_l^Y$. We assume that $\boldsymbol{\epsilon}_0^k\sim N(\mathbfit{0},\boldsymbol{\Sigma}_{Y,k})$, where $\boldsymbol{\epsilon}_0^k=(\epsilon_{0,1}^{k,1},\ldots,\epsilon_{0,1}^{k,R_1^Y},\ldots,$ $\epsilon_{0,p_Y}^{k,1},\ldots,\epsilon_{0,p_Y}^{k,R_{p_Y}^Y})^\top\in \mathbb{R}^{R^Y}$.
  
For the regression coefficients $\boldsymbol{\beta}_{0}^k(t)=(\beta_{0,1}^k(t),\ldots, \beta_{0, p_Y}^k(t))^T$ and the $p_Y\times p_X$ matrix $\boldsymbol{\beta}^k(t,s)=\left(\beta_{lj}^k(t,s)\right)_{\substack{l=1,\ldots,p_Y\\j=1,\ldots,p_X}}$ we consider the expansions \cite[Chapter 11.3]{RamsaySilverman:2006}:
 \begin{align}
 \beta_{0,l}^k(t)&=\sum_{r=1}^{R_l^Y}\Gamma_{0,l}^{k,r}\xi_{Y,r}^l(t), \quad l=1,\ldots,p_Y\label{al1}\\
\beta_{lj}^k(t,s)&= \sum_{r_1=1}^{R_l^Y}\sum_{r_2=1}^{R_j^X}\Gamma^{k,r_1r_2}_{lj}\xi_{Y,r_1}^l(t)\xi_{X,r_2}^j(s),\quad l=1,\ldots,p_Y,\quad j=1,\ldots,p_X.\label{al2}
 \end{align}
 Notice that we have 
 \begin{equation}
 \boldsymbol{\beta}^k(t,s)=\boldsymbol{\xi}_Y(t)\boldsymbol{\Gamma}^k \boldsymbol{\xi}_X(s)^\top,\quad \boldsymbol{\beta}_{0}^k(t)=\boldsymbol{\xi}_Y(t)\boldsymbol{\Gamma}_0^k,\label{eqig}
 \end{equation}
  where $\boldsymbol{\Gamma}_0^k=(\Gamma_{0,1}^{k,1},\ldots,\Gamma_{0,1}^{k,R_1^Y},\ldots,\Gamma_{0,p_Y}^{k,1},\ldots,\Gamma_{0,p_Y}^{k,R_{p_Y}^Y})^\top$ $\in \mathbb{R}^{R^Y}$ and
 \begin{equation*}
 \boldsymbol{\Gamma}^k=\begin{pmatrix}
\Gamma_{11}^{k,11}&\ldots&\Gamma^{k,1R_1^X}_{11}&\Gamma_{12}^{k,11}&\ldots&\Gamma^{k,1R_2^X}_{12}&\ldots& \Gamma^{k,11}_{1p_X}&\ldots&\Gamma^{k,1R_{p_X}^X}_{1p_X}\\
 \vdots&\ddots&\vdots&\vdots&\ddots&\vdots&\ldots&\vdots&\ddots&\vdots\\
 \Gamma_{11}^{k,R_1^Y1}&\ldots&\Gamma^{k,R_1^YR_1^X}_{11}&\Gamma_{12}^{k,R_1^Y1}&\ldots&\Gamma^{k,R_1^YR_2^X}_{12}&\ldots& \Gamma^{k,R_1^Y1}_{1p_X}&\ldots&\Gamma^{k,R_1^YR_{p_X}^X}_{1p_X}\\
  \vdots&\ddots&\vdots&\vdots&\ddots&\vdots&\ldots&\vdots&\ddots&\vdots\\
\Gamma_{p_Y1}^{k,11}&\ldots&\Gamma^{k,1R_1^X}_{p_Y1}&\Gamma_{p_Y2}^{k,11}&\ldots&\Gamma^{k,1R_2^X}_{p_Y2}&\ldots& \Gamma^{k,11}_{p_Yp_X}&\ldots&\Gamma^{k,1R_{p_X}^X}_{p_Yp_X}\\
 \vdots&\ddots&\vdots&\vdots&\ddots&\vdots&\ldots&\vdots&\ddots&\vdots\\
 \Gamma_{p_Y1}^{k,R_{p_Y}^Y1}&\ldots&\Gamma^{k,R_{p_Y}^YR_1^X}_{p_Y1}&\Gamma_{p_Y2}^{k,R_{p_Y}^Y1}&\ldots&\Gamma^{k,R_{p_Y}^YR_2^X}_{p_Y2}&\ldots& \Gamma^{k,R_{p_Y}^Y1}_{p_Yp_X}&\ldots&\Gamma^{k,R_{p_Y}^YR_{p_X}^X}_{p_Yp_X}
 \end{pmatrix}.
 \end{equation*}

 Let  ${\mathbfit{W}_X}$ be the symmetric block-diagonal $R^X\times R^X$ matrix of inner products between the basis functions:
$$
{\mathbfit{W}_X}=\int_{\mathcal{T}_X}\boldsymbol{\xi}_X(s)^\top\boldsymbol{\xi}_X(s)ds, 
$$
Using  \eqref{eqsecond}-\eqref{eqig}, for any $i=1,\ldots, n$ for which $Z_{ik}=1$  we get , 
\begin{align*}
&\mathbfit{Y}_i^\top(t)={\mathbfit{c}_{Y,i}}^\top\boldsymbol{\xi}_Y^\top (t)=\boldsymbol{\beta}_{0}^k(t)^\top+\int_{\mathcal{T}_X}\mathbfit{X}_i^\top (s)\boldsymbol{\beta}^k(t,s)^\top ds+\mathbfit{E}^k(t)^\top\\
&=\left(\boldsymbol{\Gamma}_0^k\right)^\top\boldsymbol{\xi}_Y^\top(t)+{\mathbfit{c}_{X,i}}\top\int_{\mathcal{T}_X}\boldsymbol{\xi}_X^\top (s)\boldsymbol{\xi}_X(s)ds\left(\boldsymbol{\Gamma}^k\right)^\top \boldsymbol{\xi}_Y^\top(t)+\left(\boldsymbol{\epsilon}_0^k\right)^\top
\boldsymbol{\xi}_Y^\top(t)\\
&=\left(\left(\boldsymbol{\Gamma}_0^k\right)^\top+{\mathbfit{c}_{X,i}}^\top{\mathbfit{W}_X}\left(\boldsymbol{\Gamma}^k\right)^\top +\left(\boldsymbol{\epsilon}_0^k\right)^\top\right)
\boldsymbol{\xi}_Y^\top(t),
\end{align*}
for any $t\in \mathcal{T}_Y$. Here  ${\mathbfit{c}}_{X,i}$, ${\mathbfit{c}}_{Y,i}$ are column vectors formed with the coefficients  in the $i$th row of the matrices ${\mathbfit{C}_X}$ and  ${\mathbfit{C}_Y}$ respectively.
Thus, given that $Z_{ik}=1$, we obtain the following model for  the column vector formed with the coefficients $\mathbfit{c}_{Y,i}$  in the $i$th row of the matrix $\mathbfit{C}_Y$:
\begin{equation}
\mathbfit{c}_{Y,i}=\boldsymbol{\Gamma}_0^k+\boldsymbol{\Gamma}^k\mathbfit{W}_X{\mathbfit{c}_{X,i}} +\boldsymbol{\epsilon}_0^k.\label{coeffsmat}
\end{equation}

If the curves are observed with noise, we use least square smoothing  to get the expansion for each curve \citep{RamsaySilverman:2006}.  Fourier bases are usually used for data with a repetitive pattern and B-splines functions for smooth curves \citep{SchmutzJacquesBouveyronChezeMartin:2020}. The number of basis functions depends on the data and can be chosen using cross-validation.
 %%%%%%%%%%%%%%%%%%%%%%%%%
 %%%%%%%%%%%%%%%%%%%%%%%%%%%%
 \subsection{The functional latent mixture model}
 \label{sectmod}
 %%%%%%%%%%%%%%%%%%%%%%%%%%%%%%%%
 %%%%%%%%%%%%%%%%%%%%%%%%%%%%%%%%

%%%%%%%%%%%%%%%%%%%%%%%%%%%%

We assume that for every $k\in\{1,\ldots,K\}$ the stochastic process $\mathbfit{X}$ associated with the $k$th cluster can be described in a lower dimensional subspace $\mathbb{E}^k[0,\mathcal{T}_X]\subset L^2[0,\mathcal{T}_X]$ with dimension $d_k\le R^X$ and spanned by the first $d_k$ elements of a group specific basis of functions $\{\boldsymbol{\zeta}_{X,kr}, r=1,\ldots, R^X\}$ that can be obtained from  $\{\xi_{X,r}^l, l=1,\ldots, p_X,r=1,\ldots,R^X\}$ by a linear transformation using a  MFPCA such that we have
$$
{\boldsymbol{\zeta}}_{X,kr}(t)=\sum_{j=1}^{R{^X}} q_{krj}{\boldsymbol {\xi}}_{X,j}(t),\quad r=1,\ldots,R^X,
$$
where ${\mathbfit{Q}}_k=(q_{krj})_{r,j=1,\ldots,R^X}$ is the orthogonal $R^X \times R^X$ matrix containing the coefficients of the eigenfunctions expressed in the initial basis ${\boldsymbol{\xi}}$. Using \eqref{eqsecondbis} the MFPCA scores can be obtained directly from a principal component analysis of the coefficients ${\mathbfit{C_X}}$ with a metric based on the inner products between the basis functions included in ${\mathbfit{W_X}}$. We suppose that the first $d_k$ eigenfunctions contain the main information of the MFPCA of cluster $k$ and we split
${\mathbfit Q}_k=[{\mathbfit{U}}_k,{\mathbfit{V}}_k]$ such that ${\mathbfit{U}}_k$ is of size $R^X\times d_k$, ${\mathbfit{V}}_k$ is of size $R^X\times (R^X- d_k)$ and we have
\begin{equation*}
{\mathbfit Q}_k^\top {\mathbfit Q}_k={\mathbfit I}_{R^X},\quad {\mathbfit{U}}_k^\top {\mathbfit{U}}_k={\mathbfit I}_{d_k},\quad {\mathbfit{V}}_k^\top {\mathbfit{V}}_k={\mathbfit I}_{R^X-d_k},\quad {\mathbfit{U}}_k^\top {\mathbfit{V}}_k=\mathbf{0}.
\end{equation*}
The relationship between the column vector formed with the coefficients ${\mathbfit{c}}_{X,i}$ in the $i$th row of the matrix ${\mathbfit{C}_X}$ and the score ${\boldsymbol{\delta}}_i$ is
\begin{equation*}
{\mathbfit{c}}_{X,i}={\bf{W_X}}^{-1/2}{\mathbfit{U}}_k\boldsymbol{\delta}_i+\mathbfit{e}_i,
\end{equation*}
where $\mathbfit{e}_i$ is the noise. 

We can make distribution assumptions on the scores $\boldsymbol{\delta}_i$ \citep{DelaigleHall:2010}, so, as for the model associated with the funHDDC method \citep{SchmutzJacquesBouveyronChezeMartin:2020}, we assume that independently for $i=1,\ldots, n$
 \begin{align*}
&\mathbfit{e}_i\mid Z_{ik}=1\sim N({\bf 0},\boldsymbol\Lambda_k), \text{ and } \boldsymbol{\delta}_i\mid Z_{ik}=1 \sim N({\mathbfit{m}}_k,\boldsymbol\Delta_k).
\end{align*}
 Thus 
 \begin{equation}
 \mathbfit{c}_{X,i}\mid Z_{ik}=1 \sim N({\boldsymbol{\mu}}_{X,k},\boldsymbol{\Sigma}_{X,k}).\label{distc}
 \end{equation}
Let $\phi({\mathbfit{c}}_{X,i};\boldsymbol{\mu}_{X,k},\boldsymbol{\Sigma}_{X,k})$ denotes the density for the $R_X-$variate normal distribution $N(\boldsymbol{\mu}_{X,k},\boldsymbol{\Sigma}_{X,k})$ 
\begin{align}
&\phi({\mathbfit{c}}_{X,i};\boldsymbol{\mu}_{X,k},\boldsymbol{\Sigma}_{X,k})=(2\pi)^{-R_X/2}\mid \boldsymbol{\Sigma}_{X,k}\mid ^{-1/2}\exp\biggl(-\frac{1}{2}({\mathbfit{c}}_{X,i}-\boldsymbol{\mu}_{X,k})^\top\notag\\
&\boldsymbol{\Sigma}_{X,k}^{-1}({\mathbfit{c}}_{X,i}-\boldsymbol{\mu}_{X,k})\biggl).\label{densN}
\end{align} 
Here  $\mid \boldsymbol{\Sigma}_{X,k}\mid$ denotes the determinant of $\boldsymbol{\Sigma}_{X,k}$, and
\begin{align}
& \boldsymbol{\mu}_{X,k}={\mathbfit{W}_X}^{-1/2}{\mathbfit{U}}_k{\mathbfit{m}_{k}}, \quad \boldsymbol\Sigma_{X,k}={\mathbfit{W}_X}^{-1/2}{\mathbfit{U}}_k\boldsymbol{\Delta}_k{\mathbfit{U}}_k^\top{\mathbfit{W}_X}^{-1/2}+\boldsymbol\Lambda_k,\label{dis4}
\end{align}
where the noise covariance $\boldsymbol{\Lambda}_k$ is such that the covariance ${\mathbfit{D}}_k$ of the data in the space generated by the eigenfunctions $\boldsymbol\zeta_{X,kr}$ is a diagonal matrix given by
\begin{equation}
{\mathbfit{D}}_k={\mathbfit{Q}}_k^\top {\mathbfit{W}_X}^{1/2}\boldsymbol\Sigma_{X,k}{\mathbfit{W}_X}^{1/2}{\mathbfit{Q}}_k=\text{diag}(a_{k1},\ldots, a_{kd_k}, b_k,\ldots, b_k),\label{dis5}
\end{equation}
with $a_{k1}>a_{k2}>\cdots >a_{kd_k}> b_k$. 

Given that $Z_{ik}=1$, from \eqref{coeffsmat} we have for  the column vector formed with the coefficients $\mathbfit{c}_{Y,i}$ in the $i$th row of the matrix $\mathbfit{C}_Y$:
\begin{equation*}
\mathbfit{c}_{Y,i}=\boldsymbol{\Gamma}^k_{*}{\mathbfit{c}_{X,i}^{*}}+\boldsymbol{\epsilon}_0^k,
\end{equation*}
where $\mathbfit{c}_{X,i}^{*}=\begin{pmatrix} \mathbfit{W}_X\mathbfit{c}_{X,i} \\ 1 \end{pmatrix}$ and $\boldsymbol{\Gamma}^k_{*}$ is the $R_Y\times (R_{X}+1) $ matrix $\boldsymbol{\Gamma}^k_{*}=(\boldsymbol{\Gamma}^k,\boldsymbol{\Gamma}_0^k)$.
Thus 
\begin{equation}
\mathbfit{c}_{Y,i}\mid Z_{ik}=1, \mathbfit{c}_{X,i} \sim N({\boldsymbol{\mu}}_{Y,k},\boldsymbol{\Sigma}_{Y,k}),\quad \boldsymbol{\mu}_{Y,k}=\boldsymbol{\Gamma}^k_{*}{\mathbfit{c}_{X,i}^{*}} \label{disty}
\end{equation}

Thus the joint distribution of the coefficients $(\mathbfit{c}_{Y,i}\mathbfit{c}_{X,i})$,  $i=1,\ldots, n$ arise from a parametric mixture distribution
\begin{align}
p(\mathbfit{c}_{Y,i},\mathbfit{c}_{X,i};\boldsymbol{\theta})&=\sum_{k=1}^K\pi_k p_k(\mathbfit{c}_{Y,i},\mathbfit{c}_{X,i} \mid \boldsymbol{\theta}_k), \quad \sum_{k=1}^K\pi_k=1, \label{dfun0}\\
p_k(\mathbfit{c}_{Y,i},\mathbfit{c}_{X,i} \mid \boldsymbol{\theta}_k)&=f_k(\mathbfit{c}_{X,i} \mid \boldsymbol{\theta}_k)g_k(\mathbfit{c}_{Y,i} \mid \mathbfit{c}_{X,i},\boldsymbol{\theta}_k),\label{dens}
\end{align}
where $\pi_k\in (0,1]$ are the mixing proportions, $\boldsymbol{\theta}_k=\{\boldsymbol{\mu}_{X,k}, a_{kj}, b_k, \mathbfit{q}_{kj}$, $\boldsymbol{\Sigma}_{Y,k}, \boldsymbol{\Gamma}^k_{*}\}$ and $\boldsymbol{\theta}=\bigcup_{k=1}^k (\boldsymbol{\theta}_k \cup\{\pi_k\})$,  is the set formed with the parameters. Here $f_k({\mathbfit{c}}_{X,i}\mid \boldsymbol{\theta}_k)=\phi({\mathbfit{c}}_{X,i};\boldsymbol{\mu}_{X,k},\boldsymbol{\Sigma}_{X,k})$ is given in \eqref{densN},  and $g_k({\mathbfit{c}}_{Y,i}\mid {\mathbfit{c}}_{X,i},\boldsymbol{\theta}_k)$ is the conditional density of the multivariate response ${\mathbfit{c}}_{Y,i}$ given the covariates ${\mathbfit{c}}_{X,i}$  and $Z_{ik}=1$. From \eqref{disty} we have $g_k({\mathbfit{c}}_{Y,i}\mid {\mathbfit{c}}_{X,i},\boldsymbol{\theta}_k)=\phi({\mathbfit{c}}_{Y,i};\boldsymbol{\mu}_{Y,k},\boldsymbol{\Sigma}_{Y,k})$.
 
%%%%%%%%%%%%%%%%%%%%%%%%%%%%%%%%%%C-funHDDC
As in \cite{SchmutzJacquesBouveyronChezeMartin:2020} we refer to this model for $\mathbfit{X}$ as FLM[$a_{kj},b_k,{\mathbfit{Q}}_k,d_k$] (functional latent mixture) and we consider the parsimonious sub-models:
\begin{itemize}
\item FLM[$a_{kj},b,{\mathbfit{Q}}_k,d_k$]: the parameters $b_k$ are common between the clusters
\item  FLM[$a_{k},b_k,{\mathbfit{Q}}_k,d_k$]: the first $d_k$ diagonal elements of $\mathbf{D}_k$ are common within each class
\item  FLM[$a,b_k,{\mathbfit{Q}}_k,d_k$]: the first $d_k$ diagonal elements of $\mathbfit{D}_k$ are common within each class and between the clusters
\item FLM[$a_k,b,{\mathbfit{Q}}_k,d_k$]: the parameters $b_k$ are common between the clusters and  the first $d_k$ diagonal elements of $\mathbfit{D}_k$ are common within each class
\item FLM[$a,b,{\mathbfit{Q}}_k,d_k$]: the parameters $b_k$ are common between the clusters and  the first $d_k$ diagonal elements of $\mathbfit{D}_k$ are common within each class and between the clusters
\end{itemize}
\begin{table}[h!]
\caption{Number of free parameters for the FLM models}
\label{table00}
\begin{tabular}{@{}ll|ll@{}}
\toprule
Model&Number of free parameters&Model&Number of free parameters  \\
\midrule
FLM[$a_{kj},b_k,{\mathbfit{Q}}_k,d_k$]&$\tau_1+\tau_2+2K+\sum_{k=1}^K d_k$&FLM[$a,b_k,{\mathbfit{Q}}_k,d_k$]&$\tau_1+\tau_2+2K+1$\\
FLM[$a_{kj},b,{\mathbfit{Q}}_k,d_k$]&$\tau_1+\tau_2+K+1+\sum_{k=1}^K d_k$&FLM[$a_k,b,{\mathbfit{Q}}_k,d_k$]&$\tau_1+\tau_2+2K+1$\\
FLM[$a_{k},b_k,{\mathbfit{Q}}_k,d_k$]&$\tau_1+\tau_2+3K$&FLM[$a,b,{\mathbfit{Q}}_k,d_k$]&$\tau_1+\tau_2+K+2$\\
\\
\botrule
\end{tabular}
\end{table}
Next we analyze the complexity of the FLM models. Let $\tau_1=KR_X+K-1$ be the number of parameters required for the estimation of the means $\boldsymbol{\mu}_k$ and the proportions $\pi_k$ and $\tau_2=\sum_{k=1}^K d_k[R_X-(d_k+1)/2]$ be the number of parameters required for the estimation of the matrices $\mathbfit{Q}_k$. Adding also the number of parameters required for the estimation of $b_k$, $a_{kj}$, and $d_k$ gives the total number of parameters to be estimated as included in Table \ref{table00} (see also Table 1 in \cite{ BouveyronJacques:2011}).

\begin{table}[h!]
\caption{Parsimonious models and the number of free parameters for the eigen-decomposed $\boldsymbol{\Sigma}_{Y,k} $}
\label{table0}
\begin{tabular}{@{}lll@{}}
\toprule
Model&$\boldsymbol{\Sigma}_{Y,k} $&Number of free parameters for $\boldsymbol{\Sigma}_{Y,k} $  \\
\midrule
EII&$\lambda \mathbfit{I}_{R^Y}$&1\\
VII&$\lambda_k \mathbfit{I}_{R^Y}$&K\\
EEI&$\lambda \boldsymbol{\Upsilon}$&$R^Y$\\
VEI&$\lambda_k \boldsymbol{\Upsilon}$&$K+R^Y-1$\\
EVI&$\lambda \boldsymbol{\Upsilon}_k$&$KR^Y-(K-1)$\\
VVI&$\lambda_k \boldsymbol{\Upsilon}_k$&$KR^Y$\\
EEE&$\lambda\boldsymbol{\Xi}\boldsymbol{\Upsilon}\boldsymbol{\Xi}^\top$&$R^Y(R^Y+1)/2$\\
VEE&$\lambda_k\boldsymbol{\Xi}\boldsymbol{\Upsilon}\boldsymbol{\Xi}^\top$&$R^Y(R^Y+1)/2+K-1$\\
EVE&$\lambda\boldsymbol{\Xi}\boldsymbol{\Upsilon}_k\boldsymbol{\Xi}^\top$&$R^Y(R^Y+1)/2+(K-1)(R^Y-1)$\\
EEV&$\lambda\boldsymbol{\Xi}_k\boldsymbol{\Upsilon}\boldsymbol{\Xi}_k^\top$&$KR^Y(R^Y+1)/2-(K-1)R^Y$\\
VVE&$\lambda_k\boldsymbol{\Xi}\boldsymbol{\Upsilon}_k\boldsymbol{\Xi}^\top$&$R^Y(R^Y+1)/2+(K-1)R^Y$\\
VEV&$\lambda_k\boldsymbol{\Xi}_k\boldsymbol{\Upsilon}\boldsymbol{\Xi}_k^\top$&$KR^Y(R^Y+1)/2-(K-1)(R^Y-1)$\\
EVV&$\lambda\boldsymbol{\Xi}_k\boldsymbol{\Upsilon}_k\boldsymbol{\Xi}_k^\top$&$KR^Y(R^Y+1)/2-(K-1)$\\
VVV&$\lambda_k\boldsymbol{\Xi}_k\boldsymbol{\Upsilon}_k\boldsymbol{\Xi}_k^\top$&$KR^Y(R^Y+1)/2$\\
\\
\botrule
\end{tabular}
\end{table}

As in \cite{CeleuxGovaert:1995} we consider parsimony also for the matrices $\boldsymbol{\Sigma}_{Y,k}$. An eigen-decomposition gives $\boldsymbol{\Sigma}_{Y,k} = \lambda_k\boldsymbol{\Xi}_k\boldsymbol{\Upsilon}_k\boldsymbol{\Xi}_k^\top$, where $\lambda_k=\mid\boldsymbol{\Sigma}_{Y,k} \mid^{1/R^Y}$ is a constant,$\boldsymbol{\Upsilon}_k$  is a diagonal matrix
with entries (sorted in decreasing order) proportional to the eigenvalues of
$\boldsymbol{\Sigma}_{Y,k} $ with the constraint $\mid\boldsymbol{\Upsilon}_k\mid = 1$, and $\boldsymbol{\Xi}_k$ is a $R^Y \times R^Y$ orthogonal matrix of the
eigenvectors (ordered according to the eigenvalues) of $\boldsymbol{\Sigma}_{Y,k} $,  $k = 1, \ldots K$.
We get the models in Table \ref{table0}. 

We denote a combination between a FLM[$a_{kj},b_k,{\mathbfit{Q}}_k,d_k$] model and  a VVV covariance structure  as a FLM[$a_{kj},b_k,{\mathbfit{Q}}_k,d_k$] - VVV model. We have $6 \times 14=84$ parsimonious models.  The number of parameters $\tau$ to be estimated for these models is equal with the sum between the parameters coming from the FLM model (see Table \ref{table00})  and the parameters pertaining to the model for the covariance structure (see the last column in Table \ref{table0}). 
%%%%%%%%%%%%%%%%%%%%%%%%%%%%%%%%%%T-funHDDC
%%%%%%%%%%%%%%%%%%%%%%%%%%%%%%%
\subsection{Model Identifiability}
%%%%%%%%%%%%%%%%%%%%%%%%%%%%%%%%
%%%%%%%%%%%%%%%%%%%%%%%%%%%%
Before presenting an EM algorithm for parameter estimation, it is important to study identifiabilty.   Identifiabilty of the functional model \eqref{reg1}, reduces to the identifiabilty of the model \eqref{dfun0}-\eqref{dens} in the space of the coefficients, which, roughly speaking, means that two sets of parameters $\boldsymbol{\varphi}$ and $\tilde{\boldsymbol{\varphi}}$ that do not agree after permutation cannot give the same mixture distribution. 

Identifiablity of multivariate  finite Gaussian mixture distributions is shown in \cite{YakowitzSpragins:1968}, and general conditions for identifiability of mixture of linear models  are given in \cite{Hennig:2000}.  Here we study the identifiability of the most general model FLM[$a_{kj},b_k,{\mathbfit{Q}}_k,d_k$] - VVV following the approach used in
 \cite{DangPunzoMcNicholasIngrassiaBrowne:2017}, where a sufficient identification condition is provided for the  CWM with multivariate correlated responses given in \eqref{CWM1}.

We define the parametric class of probability density functions  
 \begin{align*}
 &\mathcal{C}:=\biggl\{ p({\mathbfit{c}}_Y,{\mathbfit{c}}_X;\boldsymbol{\varphi})=\sum_{k=1}^K\pi_k \phi({\mathbfit{c}}_Y;\boldsymbol{\Gamma}^k_{*}{\mathbfit{c}_{X}^{*}},\boldsymbol{\Sigma}_{Y,k}) \phi({\mathbfit{c}}_{X};\boldsymbol{\mu}_{X,k},\boldsymbol{\Sigma}_{X,k}),\text{ with } \pi_k>0,\\
 &  \sum_{k=1}^K \pi_k=1, (\boldsymbol{\Gamma}^k_{*}, \boldsymbol{\Sigma}_{Y,k})\ne (\boldsymbol{\Gamma}^i_{*}, \boldsymbol{\Sigma}_{Y,i}), \text{ for } i\ne k, {\mathbfit{c}}_Y\in \mathbb{R}^{R_Y}, {\mathbfit{c}}_X\in \mathbb{R}^{R_X}, \mathbfit{c}_{X}^{*}=\begin{pmatrix} \mathbfit{W}_X\mathbfit{c}_{X} \\ 1 \end{pmatrix},\\
 &\boldsymbol{\varphi}=\left\{\boldsymbol{\Gamma}^k_{*}, \boldsymbol{\Sigma}_{Y,k},\boldsymbol{\mu}_{X,k},\boldsymbol{\Sigma}_{X,k}, \pi_k: k=1,\ldots,K\right\}, K\in \mathbb{N}\biggl\}
 \end{align*} 
 The next theorem proves that for  $p({\mathbfit{c}}_Y,{\mathbfit{c}}_X;\boldsymbol{\varphi})$, $p({\mathbfit{c}}_Y,{\mathbfit{c}}_X;\tilde{\boldsymbol{\varphi}})\in \mathcal{C}$,
 \begin{align*}
&p({\mathbfit{c}}_Y,{\mathbfit{c}}_X;\boldsymbol{\varphi})= \sum_{k=1}^K\pi_k \phi({\mathbfit{c}}_Y;\boldsymbol{\Gamma}^k_{*}{\mathbfit{c}_{X}^{*}},\boldsymbol{\Sigma}_{Y,k}) \phi({\mathbfit{c}}_{X};\boldsymbol{\mu}_{X,k},\boldsymbol{\Sigma}_{X,k}),\\
&p({\mathbfit{c}}_Y,{\mathbfit{c}}_X;\tilde{\boldsymbol{\varphi}})= \sum_{k=1}^{\tilde{K}}\tilde{\pi}_k \phi({\mathbfit{c}}_Y;\tilde{\boldsymbol{\Gamma}}^k_{*}{\mathbfit{c}_{X}^{*}},\tilde{\boldsymbol{\Sigma}}_{Y,k}) \phi({\mathbfit{c}}_{X};\tilde{\boldsymbol{\mu}}_{X,k},\tilde{\boldsymbol{\Sigma}}_{X,k}),
 \end{align*}
 we have $p({\mathbfit{c}}_Y,{\mathbfit{c}}_X;\boldsymbol{\varphi})=p({\mathbfit{c}}_Y,{\mathbfit{c}}_X;\tilde{\boldsymbol{\varphi}}) $ for almost all ${\mathbfit{c}}_X\in  \mathbb{R}^{R_X}$ and for all ${\mathbfit{c}}_Y\in  \mathbb{R}^{R_Y}$ if and only if $K=\tilde{K}$, and for each $k\in\{1,\ldots, K\}$ there exists $i\in\{1,\ldots, K\}$ such that $\boldsymbol{\Gamma}^k_{*}=\tilde{\boldsymbol{\Gamma}}^i_{*}$, $\boldsymbol{\Sigma}_{Y,k}=\tilde{\boldsymbol{\Sigma}}_{Y,i}$,  $\boldsymbol{\mu}_{X,k}=\tilde{\boldsymbol{\mu}}_{X,i}$, $\boldsymbol{\Sigma}_{X,k}=\tilde{\boldsymbol{\Sigma}}_{X,i}$, and $\pi_k=\tilde{\pi}_i$.
 \begin{theorem}
 Suppose that there exists a set $\mathcal{X}\in\mathbb{R}^{R_X}$ having probability equal to one according to the $R_X$-variate Gaussian distribution such that the mixture of regression models
 $$
 \sum_{k=1}^K\phi({\mathbfit{c}}_Y;\boldsymbol{\Gamma}^k_{*}{\mathbfit{c}}_{X}^{*},\boldsymbol{\Sigma}_{Y,k})\alpha_k({\mathbfit{c}}_X),\quad {\mathbfit{c}}_Y\in  \mathbb{R}^{R_Y},
 $$
 is identifiable for each fixed ${\mathbfit{c}}_X\in \mathcal{X} $, where  $\alpha_k({\mathbfit{c}}_X)>0$, $k=1,\ldots, K$, $\sum_{k=1}^K\alpha_k({\mathbfit{c}}_X)=1$ for each ${\mathbfit{c}}_X\in \mathcal{X} $. Then the class $\mathcal{C}$ is identifiable in $\mathcal{X}\times \mathbb{R}^{R_Y}$.
 \end{theorem}
 \begin{proof}
 The proof is similar the proof of Theorem 1 in \cite{DangPunzoMcNicholasIngrassiaBrowne:2017}.
 \end{proof}
 
%%%%%%%%%%%%%%%%%%%%%%%%%%%%%%%%
%%%%%%%%%%%%%%%%%%%%%%%%%%%%%%%%%%%%%%

%%%%%%%%%%%%%%%%%%%%%%%%%%%%

%%%%%%%%%%%%%%%%%%%%%%%%%%%%%%%%
%%%%%%%%%%%%%%%%%%%%%%%%%%%%
\section{Learning the FLM[$a_{kj},b_k,{\mathbfit{Q}}_k,d_k$] - VVV model}
\label{section3}
 %%%%%%%%%%%%%%%%%%%%%%%%%%%%%%%%%%
 %%%%%%%%%%%%%%%%%%%%%%%%%%%

%%%%%%%%%%%%%%%%%%
%%%%%%%%%%%%%%%%%%%%%%%%%%%%%%%%%%
%%%%%%%%%%%%%%%%%%%%%%%%%%%%%%%%%
%%%%%%%%%%%%%%%%%%%%%%%%%%%%%%
%%%%%%%%%%%%%%%%%%%%%%%%%%%%%%
%%%%%%%%%%%%%%%%%%%%%%%%%%
%%%%%%%%%%%%%%%%%%%%%%%%%%%%%%
%%%%%%%%%%%%%%%%%%%%%%%%%%%%%%%%
To fit the models, we use the  expectation-maximization (EM) algorithm \citep{DempsterLairdRubin:1977}.  The algorithm consists of successive iterations of the expectation (E) and the maximization (M) steps until convergence is achieved.  The clusters' labels ${\mathbfit{Z}_i}$ are the missing data, so the complete data are given by $\{\mathbfit{c}_{Y,i}, \mathbfit{c}_{X,i}, z_{ik}, i=1,\ldots, n, k=1,\ldots,K\}$.  The current estimates of the parameters are used in the  E step to compute the conditional expectation of the complete log-likelihood. In the M step  the estimates of the parameters are updated with the values that maximize the expected complete log-likelihood. Next we present the EM algorithm for the most general model   FLM[$a_{kj},b_k,{\mathbfit{Q}}_k,d_k$] - VVV model.  The main steps of the proposed EM algorithm are summarized in Algorithm \ref{algo1}.

 %%%%%%%%%%%%%%%%%%%%%%%%%%%%%%%%
 \begin{proposition}
 \label{proplik}
 The complete data log-likelihood of the observed curves under the FLM[$a_{kj},b_k,\mathbfit{Q}_k,d_k$] - VVV model  for $\mathbfit{X}$  can be written as
 \begin{equation}
l_c(\boldsymbol{\theta})=l_{1c}(\pi)+l_{2c}(\boldsymbol{\vartheta}_X)+l_{3c}(\boldsymbol{\vartheta}_Y)\label{lik2}
\end{equation}
where
\begin{align}
&l_{1c}(\pi)=\sum_{i=1}^n\sum_{k=1}^K z_{ik}\log(\pi_k),\label{l1c2}\\
&l_{2c}(\boldsymbol{\vartheta}_X)==-\frac{nR_X\log(2\pi)}{2}+\frac{n}{2}\log(\mid \mathbfit{W}_X\mid)-\frac{1}{2}\sum_{k=1}^K n_k\sum_{l=1}^{d_k}\log(a_{kl})\nonumber\\
&-\frac{1}{2}\sum_{k=1}^K n_k\sum_{l=d_k+1}^{R_X}\log(b_{k})-\frac{1}{2}\sum_{k=1}^K   \biggl(\sum_{l=1}^{d_k}\frac{\mathbfit{q}_{kl}^\top \mathbfit{W}_X^{1/2}\mathbfit{S}_{X,k}\mathbfit{W}_X^{1/2}\mathbfit{q}_{kl}}{a_{kl}}\nonumber\\
&+\sum_{l=d_k+1}^{R}\frac{\mathbfit{q}_{kl}^\top \mathbfit{W}_X^{1/2}\mathbfit{S}_{X,k}\mathbfit{W}_X^{1/2}\mathbfit{q}_{kl}}{b_{k}}\biggl)\label{l3c2}\\
&l_{3c}(\boldsymbol{\vartheta}_Y)=-\frac{nR_Y\log(2\pi)}{2}-\frac{n}{2}\sum_{k=1}^K n_k\log\mid \boldsymbol{\Sigma}_{Y,k}\mid -\frac{1}{2}\sum_{i=1}^n\sum_{k=1}^K z_{ik}\biggl(\mathbfit{c}_{Y,i}^\top\boldsymbol{\Sigma}_{Y,k}^{-1}\mathbfit{c}_{Y,i}\nonumber\\
&-\mathbfit{c}_{Y,i}^\top\boldsymbol{\Sigma}_{Y,k}^{-1}\boldsymbol{\Gamma}^k_{*}{\mathbfit{c}_{X,i}^{*}}-({\mathbfit{c}_{X,i}^{*}})^\top\left(\boldsymbol{\Gamma}^k_{*}\right)^\top \boldsymbol{\Sigma}_{Y,k}^{-1}\mathbfit{c}_{Y,i}+({\mathbfit{c}_{X,i}^{*}})^\top\left(\boldsymbol{\Gamma}^k_{*}\right)^\top\boldsymbol{\Sigma}_{Y,k}^{-1}\boldsymbol{\Gamma}^k_{*}{\mathbfit{c}_{X,i}^{*}}\biggl)\label{l2c2}
\end{align}
where  $\boldsymbol{\vartheta}_X=\{\boldsymbol{\mu}_{X,k}, a_{kj}, b_k, \mathbfit{q}_{kj}\}$, $\boldsymbol{\vartheta}_Y=\{\boldsymbol{\Sigma}_{Y,k}, \boldsymbol{\Gamma}^k_{*} \}$, $k=1,\ldots, K$, $j=1,\ldots, d_k$, with $\mathbfit{q}_{kj}$ the $j$th column of $\mathbfit{Q}_k$, $n_k=\sum_{i=1}^n z_{ik}$, and $\mathbfit{S}_{X,k}$ is defined by
\begin{equation}
\mathbfit{S}_{X,k}:=\sum_{i=1}^n z_{ik}(\mathbfit{c}_{X,i}-\boldsymbol{\mu}_{X,k})(\mathbfit{c}_{X,i}-\boldsymbol{\mu}_{X,k})^\top. \label{defS}
\end{equation}
 \end{proposition}
 \begin{proof}
 The proof is included in  Appendix \ref{secA1}.
 \end{proof}
 %%%%%%%%%%%%%%%%%%%%%%%%%%%%%%%

%%%%%%%%%%%%%%%%%%%%%%%%%
%%%%%%%%%%%%%%%%%%%%%%%%%%%%
%%%%%%%%%%%%%%%%%%%%%%%%%
\subsubsection{The E-step}
At the $m$th iteration of the EM algorithm we calculate $E[l_c(\boldsymbol{\theta}^{(m-1)})\mid \mathbfit{c}_{Y,1},\mathbfit{c}_{X,1}\ldots,\mathbfit{c}_{Y,n},\mathbfit{c}_{X,n},\boldsymbol{\theta}^{(m-1)}]$, given the current values of the parameters $\boldsymbol{\theta}^{(m-1)}$.  This reduces to the calculation of 
 $E[Z_{ik}\mid \mathbfit{c}_{Y,1},\mathbfit{c}_{X,1}\ldots,\mathbfit{c}_{Y,n},\mathbfit{c}_{X,n},\boldsymbol{\theta}^{(m-1)}]$.
 %%%%%%%%%%%%%%%%%
 \begin{proposition}
 \label{propEM}
Let us denote
\begin{align}
&H_k(\mathbfit{c}_{Y,i},\mathbfit{c}_{X,i}\mid \boldsymbol{\theta}_k):=-2\log (\pi_k)+\sum_{j=1}^{d_k} \log (a_{kj})+(R_X-d_k)\log (b_k)\notag\\
&+\log(\mid\boldsymbol{\Sigma}_{Y,k}\mid)+\delta(\mathbfit{c}_{X,i};\boldsymbol{\mu}_{X,k},\mathbfit{Q}_k,a,b,d_k)+({\mathbfit{c}}_{Y,i}-\boldsymbol{\Gamma}^k_{*}{\mathbfit{c}_{X,i}^{*}} )^\top\boldsymbol{\Sigma}_{Y,k}^{-1}\notag\\
&({\mathbfit{c}}_{Y,i}-\boldsymbol{\Gamma}^k_{*}{\mathbfit{c}_{X,i}^{*}} )\label{Hk}\\
&\delta(\mathbfit{c}_{X,i};\boldsymbol{\mu}_{X,k},\mathbfit{Q}_k,a,b,d_k):=\sum_{l=1}^{d_k}\frac{\mathbfit{q}_{kl}^\top\mathbfit{W}_X^{1/2}(\mathbfit{c}_{X,i}-\boldsymbol{\mu}_{X,k})(\mathbfit{c}_{X,i}-\boldsymbol{\mu}_{X,k})^\top\mathbfit{W}_X^{1/2}\mathbfit{q}_{kl}}{a_{kl}}\nonumber\\
&+\sum_{l=d_k+1}^{p}\frac{\mathbfit{q}_{kl}^\top\mathbfit{W}_X^{1/2}(\mathbfit{c}_{X,i}-\boldsymbol{\mu}_{X,k})(\mathbfit{c}_{X,i}-\boldsymbol{\mu}_{X,k})^\top\mathbfit{W}_X^{1/2}\mathbfit{q}_{kl}}{b_k}\label{newdelta}
\end{align}
We have
\begin{align}
&t_{ik}^{(m)}:=E[Z_{ik}\mid \mathbfit{c}_{Y,1},\mathbfit{c}_{X,1}\ldots,\mathbfit{c}_{Y,n},\mathbfit{c}_{X,n},\boldsymbol{\theta}^{(m-1)}]=\frac{\pi_k p_k\left(\mathbfit{c}_{Y,i},\mathbfit{c}_{X,i} \mid \boldsymbol{\theta}_k^{(m-1)}\right)}{\sum_{l=1}^K\pi_l p_l\left(\mathbfit{c}_{Y,i},\mathbfit{c}_{X,i} \mid \boldsymbol{\theta}_l^{(m-1)}\right)}\notag\\
&=\frac{1}{\sum_{l=1}^K\exp \left(\frac{1}{2}\left(H_k\left(\mathbfit{c}_{Y,i},\mathbfit{c}_{X,i}\mid \boldsymbol{\theta}_k^{(m-1)}\right)-H_l\left(\mathbfit{c}_{Y,i},\mathbfit{c}_{X,i}\mid \boldsymbol{\theta}_l^{(m-1)}\right)\right)\right)},\label{newt}
\end{align}
\end{proposition}
\begin{proof}
 The proof is included in  Appendix \ref{secA2}.
 \end{proof}
%%%%%%%%%%%%%%%%%%%%%%%%%%%%%%%
%%%%%%%%%%%%%%%%%%%%%%%%%%%%%
Based on the current values of the parameters $\boldsymbol{\theta}^{(m-1)}$ the log-likelihood is given by
\begin{align*}
\L^{(m-1)}&=\log\left(\prod_{i=1}^n p\left(\mathbfit{c}_{Y,i},\mathbfit{c}_{X,i};\boldsymbol{\theta}^{(m-1)}\right) \right)\\
&=\sum_{i=1}^n\log\left(\sum_{k=1}^K\pi_k^{(m-1)} p_k\left(\mathbfit{c}_{Y,i},\mathbfit{c}_{X,i} \mid \boldsymbol{\theta}_k^{(m-1)}\right)\right)
\end{align*}
%%%%%%%%%%%%%%%%%%%%%
%%%%%%%%%%%%%%%%%%%%%%
%%%%%%%%%%%%%%%%%%%%%
\subsubsection{The M-step}
In the M-step at the $m$th  iteration of the EM algorithm we estimate the parameters  by maximizing the conditional expectation of the complete data log likelihood  $Q(\boldsymbol{\theta}\mid \boldsymbol{\theta}^{(m-1)}):=E[\log(l_c(\boldsymbol{\theta}^{(m-1)}))\mid \mathbfit{c}_{Y,1},\mathbfit{c}_{X,1},\ldots,\mathbfit{c}_{Y,n},\mathbfit{c}_{X,n},\boldsymbol{\theta}^{(m-1)}]$.
%%%%%%%%%%%%%%%%%%%%%%%%
\begin{proposition}
\label{propM}
Let 
\begin{align}
\mathbfit{S}_{X,k}^{(m)}&=\frac{\sum_{i=1}^n t_{ik}^{(m)}(\mathbfit{c}_{X,i}-\boldsymbol{\mu}_{X,k}^{(m)})(\mathbfit{c}_{X,i}-\boldsymbol{\mu}_{X,k}^{(m)})^\top}{n_k^{(m)}}, \quad n_k^{(m)}=\sum_{i=1}^n t_{ik}^{(m)}.\label{sig}
\end{align}
For the model FLM[$a_{kj},b_k,{\mathbfit{Q}}_k,d_k$]-VVV  we have the following updates for the parameters, $k=1,\ldots, K$
\begin{align}
\pi_k^{(m)}&=\frac{\sum_{i=1}^n t_{ik}^{(m)}}{n}=\frac{n_k^{(m)}}{n}, \label{tpi}\\
\boldsymbol{\mu}_{X,k}^{(m)}&=\frac{\sum_{i=1}^n t_{ik}^{(m)}\mathbfit{c}_{X,i} }{\sum_{i=1}^n t_{ik}^{(m)}}\label{tmu}
\end{align}
\begin{align}
&(\boldsymbol{\Gamma}^k_{*})^{(m)}=\left(\sum_{i=1}^n t_{ik}^{(m)}\mathbfit{c}_{Y,i}({\mathbfit{c}_{X,i}^{*}})^\top\right)\left(\sum_{i=1}^n t_{ik}^{(m)}{\mathbfit{c}_{X,i}^{*}}({\mathbfit{c}_{X,i}^{*}})^\top\right)^{-1},\\
&\boldsymbol{\Sigma}_{Y,k}^{(m)}=\frac{\sum_{i=1}^n t_{ik}^{(m)}({\mathbfit{c}}_{Y,i}-(\boldsymbol{\Gamma}^k_{*})^{(m)}{\mathbfit{c}_{X,i}^{*}} )({\mathbfit{c}}_{Y,i}-(\boldsymbol{\Gamma}^k_{*})^{(m)}{\mathbfit{c}_{X,i}^{*}} )^\top}{n_{k}^{(m)}}.
\end{align}
\begin{itemize}
\item ${\mathbfit{q}}_{kj}^{(m)}$, $k=1,\ldots, K, j=1,\ldots, d_k$ are updated as the eigenfunctions associated with the $d_k$ largest  eigenvalues of $\mathbfit{W}_X^{1/2}\mathbfit{S}_{X,k}^{(m)}\mathbfit{W}_X^{1/2}$;
\item $a_{kj}^{(m)}$,  $k=1,\ldots, K, j=1,\ldots, d_k$ are updated by the $d_k$ largest  eigenvalues of $\mathbfit{W}^{1/2}\mathbfit{S}_{X,k}^{(m)}\mathbfit{W}_X^{1/2}$;
\item $b_k^{(m)}$, $k=1,\ldots, K$ are updated by
\begin{equation}
b_k^{(m)}=\frac{1}{R_X-d_k}\left(\text{trace}\left(\mathbfit{W}_X^{1/2}\mathbfit{S}_{X,k}^{(m)}\mathbfit{W}_X^{1/2}\right)-\sum_{j=1}^{d_k} a_{kj}^{(m)}\right).\label{tb1}
\end{equation}
\end{itemize}
\end{proposition}
\begin{proof}
 The proof is included in  Appendix \ref{secA3}.
 \end{proof}
For the other parsimonious models the updates in the M-step are different only for the covariance matrix $\boldsymbol{\Sigma}_{Y,k}^{(m)}$ and $a_{kj}^{(m)}$, $b_k^{(m)}$, $k=1,\ldots, K, j=1,\ldots, d_k$.
For the  models in Table \ref{table0} for the covariance matrix $\boldsymbol{\Sigma}_{Y,k}^{(m)}$ the updates are similar with those of the Gaussian parsimonious
clustering models in \cite{CeleuxGovaert:1995}. For the simplified FLM models we have:
\begin{itemize}
\item FLM[$a_{kj},b,{\mathbfit{Q}}_k,d_k$] for $\mathbfit{X}$: the estimator of $b$ is
\begin{equation}
b^{(m)}=\frac{\text{trace}\left(\sum_{k=1}^K\pi_k^{(m)}\mathbfit{W}_X^{1/2}\mathbfit{S}_{X,k}^{(m)}\mathbfit{W}_X^{1/2}\right)-\sum_{k=1}^K\pi_k^{(m)}\sum_{j=1}^{d_k} a_{kj}^{(m)}}{R_X-\sum_{k=1}^K\pi_k^{(m)} d_k}\label{tb}
\end{equation}
\item  FLM[$a_{k},b_k,{\mathbfit{Q}}_k,d_k$] for $\mathbfit{X}$: the estimator of $a_k$ is
\begin{equation}
a_k^{(m)}=\frac{\sum_{j=1}^{d_k} a_{kj}^{(m)}}{d_k}\label{tak}
\end{equation}
 and the estimator of $b_k$ is given by \eqref{tb1}.
\item  FLM[$a,b_k,{\mathbfit{Q}}_k,d_k$] for $\mathbfit{X}$: the estimator of $a$ is
\begin{equation}
a^{(m)}=\frac{\sum_{k=1}^K\pi_k^{(m)}\sum_{j=1}^{d_k} a_{kj}^{(m)}}{\sum_{k=1}^K\pi_k^{(m)} d_k}\label{ta}
\end{equation}
 and the estimator of $b_k$ is given by \eqref{tb1}.
\item FLM[$a_k,b,{\mathbfit{Q}}_k,d_k$] for $\mathbfit{X}$: the estimator of $a_k$ is given by \eqref{tak} and the estimator of $b$ is given by \eqref{tb}.
\item FLM[$a,b,{\mathbfit{Q}}_k,d_k$] for $\mathbfit{X}$:  the estimator of $a$ is given by \eqref{ta} and the estimator of $b$ is given by \eqref{tb}.
\end{itemize}
%%%%%%%%%%%%%%%%%%%%%%%%%%%%%%%%%%%%%
%%%%%%%%%%%%%%%%%%%%%%%%%%%%%%%%%%%%%%%%%%
\begin{algorithm}[h!]
\caption{EM algorithm for funWeightClust }\label{algo1}
\begin{algorithmic}[1]
\Require  data $(\mathbfit{y}_1,\mathbfit{x}_1), \ldots, (\mathbfit{y}_n,\mathbfit{x}_n)$, range for number of clusters $K=K_1,\ldots, K_G$, funWeightClust models $set_{models}$, initialization method $init \in \{{\it{kmeans}}, random\}$, number of repetitions for initialization $n_{rep}$, threshold for Cattell scree-test $\epsilon$, maximum number of iterations $max_{iter}$, precision $\epsilon_1$ 
\State Compute $W_X$ and $W_X^{1/2}$
\For{$K=K_1$ to $K_G$} (this can be performed in parallel)
	\For{$r=1$ to $n_{rep}$} (this can be performed in parallel)
			\State Initialization
					\State $\quad$	Apply method {\it init} to calculate $t_{ik}^{(0)}$, $i=1,\ldots, n$, $k=1,\ldots K$
			\ForAll{$model \in set_{models}$}
			\State $m \leftarrow 1$
			\State  $test \leftarrow \epsilon_1+1$
			\While{$m \le max_{iter}$ and $test \ge \epsilon_1$}
			\State M-step
				\State $\quad$Compute $d_k^{(m)}$ using Cattell scree test with threshold $\epsilon$, 					\State $\quad k=1,\ldots, K$
				\State $\quad$Compute $\boldsymbol{\theta}^{(m)}=\{\pi_k^{(m)},\boldsymbol{\mu}_{X,k}^{(m)}, a_{kj}^{(m)}, b_k^{(m)}, \mathbfit{q}_{kj}^{(m)}, \boldsymbol{\Sigma}_{Y,k}^{(m)}, \left(\boldsymbol{\Gamma}^k_{*}\right)^{(m)} \}$, 	
				\State $\quad$ $k=1,\ldots, K$, $j=1,\ldots, d_k^{(m)}$
			\State E-step
			\State $\quad$Compute $t_{ik}^{(m)}$, $i=1,\ldots, n$, $k=1,\ldots K$
			\State $\quad$Compute the log likelihood $L^{(m)}$
			\If{$m=2$ }
            \State $test \leftarrow L^{(m)}-L^{(m-1)}$
        \EndIf 
        	\If{$m>2$}
            \State Compute  Aitken acceleration $A\leftarrow\frac{L^{(m)}-L^{(m-1)}}{L^{(m-1)}-L^{(m-2)}}$
            \State Compute the asymptotic estimate  $AL\leftarrow L^{(m-1)}+\frac{L^{(m)}-L^{(m-1)}}{1-A}$
            \State Compute $test \leftarrow \mid AL-L^{(m-1)}\mid $
            \EndIf     
			\EndWhile
			\State  Compute BIC
			\State  Assign a cluster to each observation $(\mathbfit{y}_i,\mathbfit{x}_i)$, $i=1,\ldots, n$  using MAP
			\EndFor
	\EndFor
	\State Choose the model with the maximum BIC if the number of clusters is $K$ 
\EndFor
\State Choose  the number of clusters $K$ that gives the largest BIC
\Ensure the best number of clusters, the best model,  the assigned clusters
\end{algorithmic}
\end{algorithm}

%%%%%%%%%%%%%%%%%%%%%%%%%%%%%%%%%%%%%
%%%%%%%%%%%%%%%%%%%%%%%%%%%%%%%%%%%%%%%%%%%

%%%%%%%%%%%%%%%%%%%%%%%%%%
%%%%%%%%%%%%%%%%%%%%%%%%%%%%%%%%%%%
\subsubsection{Initialization}
To start the EM algorithm, we need initial values $t_{ik}^{(0)}$. As for funHDDC  \citep{SchmutzJacquesBouveyronChezeMartin:2020}, we have implemented a random initialization and an  initialization with the {\it{kmeans}} method  available in the {\it{stats}} package in R. For random initialization, the values of $t_{ik}^{(0)}$ correspond to a partition randomly sampled using a multinomial
distribution with uniform probabilities. For the {\it{kmeans}} strategy, the values of $t_{ik}^{(0)}$ correspond to the partition obtained applying the {\it{kmeans}} method to the data set formed by the combining  the coefficients  $\mathbfit{C_X}$, $\mathbfit{C_Y}$. 
 
To prevent the convergence of the EM algorithm to a local maximum,  we execute the algorithm with different initialization values for $t_{ik}^{(0)}$, and we keep the best result given by the EM algorithm using the Bayesian information criterion (BIC; \citealp{Schwarz:1978}) defined by 
\begin{equation}
BIC=L^{(m_f)}-\frac{\tau}{2}\log n,\label{bicm}
\end{equation}
where $\tau$ is the overall number of the free parameters, $n$ is the number of observations, $L^{(m_f)}$ is the maximum log-likelihood value, and $m_f$ is the last iteration of the algorithm before convergence. 
For the numerical experiments in Section \ref{section4} we consider the number of initializations to be at least 20.
%%%%%%%%%%%%%%%%%%%%%%%%%%%%%%%%%%
%%%%%%%%%%%%%%%%%%%%%%%%%%%%%%%%%%%%%%%%%
\subsubsection{ Estimation of the hyper-parameters, convergence criterion, and the classification step}
The number of clusters $K$ and the parsimonious model are selected by maximizing the the BIC given in \eqref{bicm}.  The group specific dimension $d_k$ is selected through the Cattell scree-test by comparison of the difference between eigenvalues with a given threshold $\epsilon$   \citep{BouveyronJacques:2011}. In \cite{Amovin-AssagbaGannazJacques:2022} a grid search is applied and $d_k$ is chosen as the positive integer from the grid that corresponds to the maximum value of the BIC. Since the grid search requires substantially more time and we have not obtained a major improvement using the grid search,  in Section \ref{section4} we present the results obtained with the Cattell scree-test.

To avoid spurious clusters, which are a known problem for mixture models \citep{McLachlanPeel:2000}, we proceed as in \cite{DangPunzoMcNicholasIngrassiaBrowne:2017} and we remove the models that have a matrix $\boldsymbol{\Sigma}_{Y,k}^{(m_f)}$ for which at least one eigenvalue  is less than $10^{-20}$. In the simulations presented in section \ref{simu} this procedure has the effect of disregarding all models with spurious clusters and obtaining more accurate estimations for the number of clusters $K$.

We consider that the EM algorithm has converged if a maximum number of iterations is reached or $\mid L^{(m+2)}_{\infty}-L^{(m+1)}\mid<\epsilon_1$ \citep{McNicholasMurphyMcDaidFrost:2010}, where $L^{(m+1)}$ is the log-likelihood value at iteration $m+1$, and $L^{(m+2)}_{\infty}$ is the asymptotic estimate of log-likelihood  at iteration $m+2$ \citep{AndrewsMcNicholasSubedi:2011} defined as
\begin{equation*}
L^{(m+2)}_{\infty}=L^{(m+1)}+\frac{L^{(m+2)}-L^{(m+1)}}{1-a^{(m+1)}}.
\end{equation*}
Here $a^{(m+1)}$ is the Aitken acceleration \citep{Aitken:1927} at iteration $m+1$:
\begin{equation*}
a^{(m+1)}=\frac{L^{(m+2)}-L^{(m+1)}}{L^{(m+1)}-L^{(m)}}
\end{equation*}
We choose 200 for the maximum number of iterations and  the precision $\epsilon_1=10^{-6}$.

We determine the clusters using the maximum {\it{a posteriori}} (MAP) rule: an observation $(\mathbfit{c}_{Y,i},\mathbfit{c}_{X,i})$ is assigned to the cluster $k\in\{1,\ldots, K\}$ with the largest $t_{ik}^{(m_f)}$, where $m_f$ is the last iteration of the  EM algorithm before convergence.
%%%%%%%%%%%%%%%%%%%%%%%%%%%

%%%%%%%%%%%%%%%%%%%%%%%%%%%%%%%
%%%%%%%%%%%%%%%%%%%%%%%%%%%%%%%%
%%%%%%%%%%%%%%%%%%%%%%%%%%%%
%%%%%%%%%%%%%%%%%%%%%%%%%%%%%%%%
%%%%%%%%%%%%%%%%%%%%%%%%%%%%%%%%%%%%%%

%%%%%%%%%%%%%%%%%%%%%%%%%%%%

%%%%%%%%%%%%%%%%%%%%%%%%%%%%%%%%
%%%%%%%%%%%%%%%%%%%%%%%%%%%%
  
%%%%%%%%%%%%%%%%%%%%%%%%%%%%%
%%%%%%%%%%%%%%%%%%%%%%%%%%%%%%%%%%%%%%
\section{Experiments and results}
\label{section4}
We apply the proposed clustering method to simulated data, the Adelaide electricity demand data (available in the {\it fds} package in R), and Edmonton traffic data. For some of these clustering examples, the true classifications are known and the Adjusted Rand Index (ARI; \citealp{HubertArabie:1985}) is used to
measure the accuracy of the classification. The expected value of the adjusted Rand index is 0,
and for a perfect classification its value is 1.  Negative values for ARI indicates that the classification is worse than would be expected by random assignment. We compare funWeightClust with methods for clustering functional data ({\it funHDDC} R package) and methods for clustering multivariate data ({\it flexmix} and {\it mclust} R packages). When we apply the methods for clustering multivariate data we use the two approaches mentioned at the beginning of section \ref{sectlit}:  a raw-data clustering  and a two-step method.
%%%%%%%%%%%%%%%%%%%%%%%%%%%%%%%%%
%%%%%%%%%%%%%%%%%%%%%%%%%%%%%%%%%%%%%%%
%%%%%%%%%%%%%%%%%%%%%%%%%%%%%%
\subsection{Simulation studies on synthetic data sets}
\label{simu}
We simulate 600 pairs of curves according to the  FLM[$a_{kj},b_k,{\mathbfit{Q}}_k,d_k$]$\times VII$ model. The number of clusters is 2 and the mixing proportions are $\pi_1=\pi_2=1/2.$ We consider two scenarios. For the first scenario for the covariates curves $X_i$ the parameters are
\begin{enumerate}
\item[] Group 1: $d_1=2$, $\mathbfit{a}_1=(6730.074, 1641.839)^\top$, $b_1=70.57964$, $\boldsymbol{\mu}_{X,1}=(1459.420, 1297.329, 883.6936, 1052.785, 1167.558, 1183.825)^\top$
\item[] Group 2: $d_2=1$, $a_2=95464.836$, $b_2=2351.284$, $\boldsymbol{\mu}_{X,2}=(1555.634, 1450.803, 867.3406, 1429.287, 1528.500, 1517.618)^\top$
\end{enumerate}
where $d_k$ is the intrinsic dimension of the subgroups, $\boldsymbol{\mu}_{X,k}$ is the mean vector of size 6, $\mathbfit{a}_k$ includes the  values of the $d_k$-first diagonal elements of $\mathbfit{D}$, and $b_k$ the value of the last $6-d_k$- elements, $k=1,2$. Curves as smoothed using 6 cubic B-spline basis functions. 
For the response curves $Y_i$ the coefficients are generated using the coefficients of the curves  $X_i$ and equation \eqref{coeffsmat} with the values of regression coefficients $\boldsymbol{\Gamma}^1$, $\boldsymbol{\Gamma}^2$, $\boldsymbol{\Gamma}_0^1$, $\boldsymbol{\Gamma}_0^2$ given in Appendix \ref{apsimu},  
and $\boldsymbol{\Sigma}_{Y,1}=434.6492 $ $\mathbfit{I}_6$, $\boldsymbol{\Sigma}_{Y,2}=1014.901$ $\mathbfit{I}_6$, where $\mathbfit{I}_6$ is the 6th dimensional identity matrix. Curves as smoothed using 6 cubic B-spline basis functions. 

\begin{figure}[h!]
\begin{center}
\includegraphics[width=12cm,height=8cm]{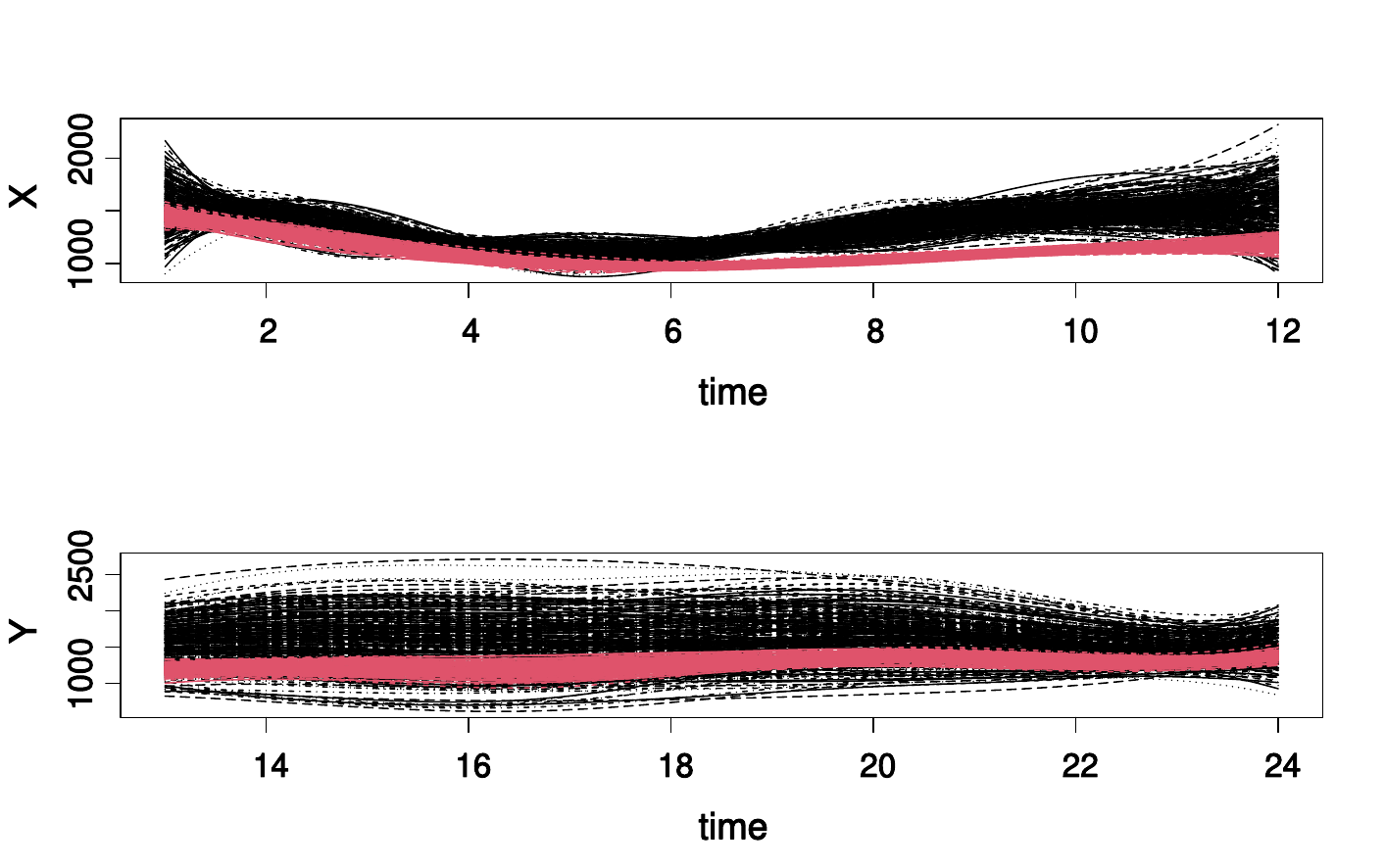}
\end{center}
  \caption{Smooth data simulated  according to scenario 1  colored by group for one simulation.}
\label{fig1}
\end{figure}

\begin{figure}[h!]
\begin{center}
\includegraphics[width=12cm,height=8cm]{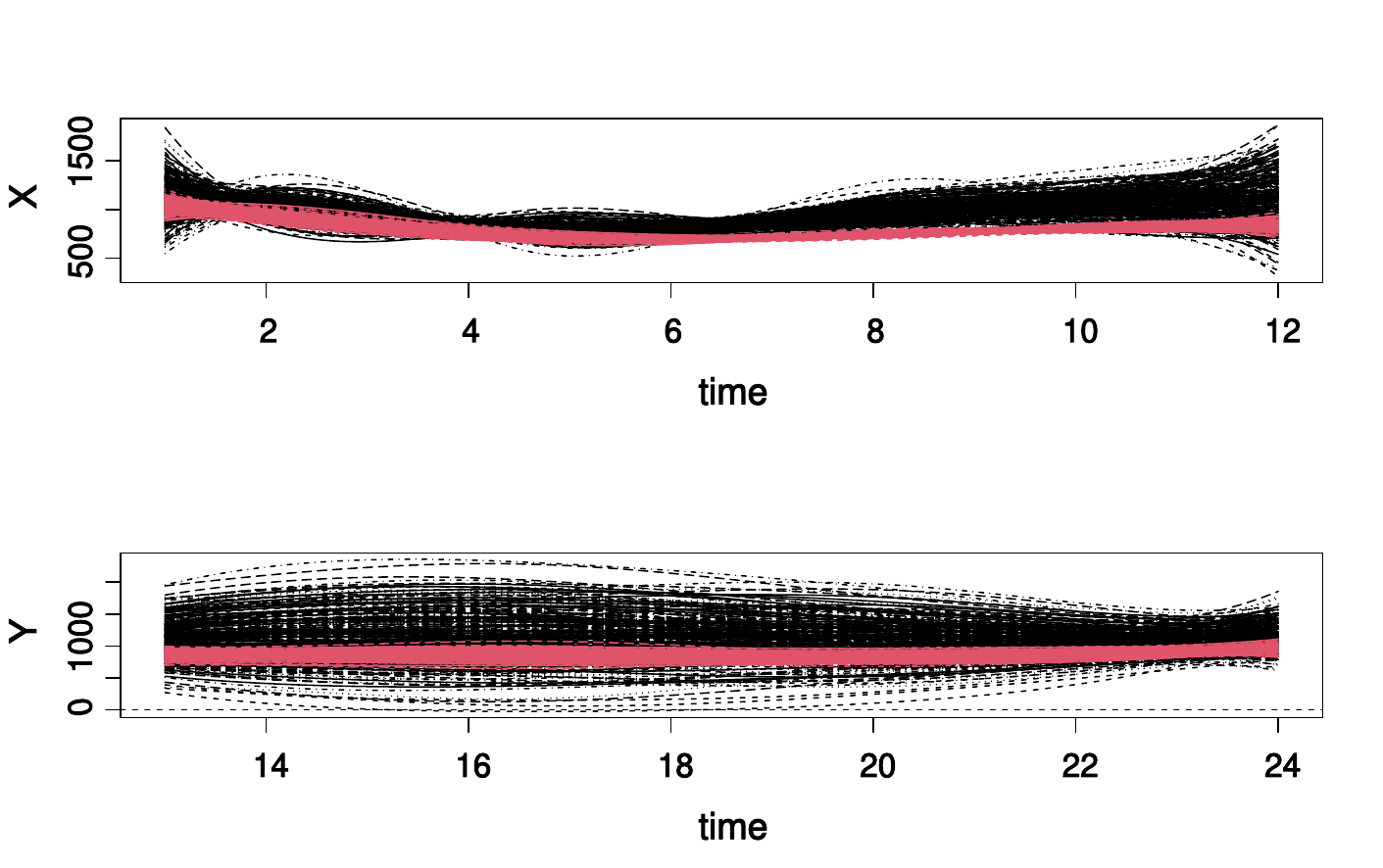}
\end{center}
  \caption{Smooth data simulated  according to  scenario 2  colored by group for one simulation.}
\label{fig2}
\end{figure}

We can notice in Figure \ref{fig1} that while the $Y_i$ curves from the two clusters overlap, there is a visible separation between the clusters for the $X_i$ curves. We also simulate the more difficult scenario illustrated in Figure \ref{fig2}. The parameters are the same as before except the values for $\boldsymbol{\mu}_{X,k}$, $k=1,2$:
\begin{enumerate}
\item[] Group 1:  $\boldsymbol{\mu}_{X,1}=(1042.4431,  926.6636 , 631.2097,  751.9895 , 833.9701,  845.5891)^\top$
\item[] Group 2:  $\boldsymbol{\mu}_{X,2}=(1111.167, 1036.288,  619.529, 1020.919, 1091.786, 1084.013 )^\top$
\end{enumerate}

  
   For both scenarios we repeat the simulation 100 times. We use ARI to evaluate the clustering done with funWeightClust and funHDDC. We apply funHDDC for the curves obtained by combining the $X_i:[0,12]\rightarrow \mathbb{R}$ and $Y_i:[12,24]\rightarrow \mathbb{R}$ curves in one curve over the time interval $[0,24]$. Alternatively, we consider the pairs of curves $(X_i,Y_i)$ as two-dimensional functional data when we apply funHDDC. We run  funWeightClust and funHDDC for $K=2$ with all sub-models, and the best solution in terms of the highest BIC value for all those sub-models is returned. The initialization is done with the {{\it{kmeans}}} method  with 50 repetitions, and the maximum number of iterations is 200 for the stopping criterion. We use $\epsilon \in\{ 0.005, 0.01, 0.2\}$ in the Cattell test. We denote the method applied to the combined curves by funHDDC, and we denote funHDDC applied to the two-dimensional curves $(X_i,Y_i)$ by 2D-funHDDC.
   
From the results included in Table \ref{table1} we notice that funWeightClust outperforms funHDDC and  2D-funHDDC. For scenario 1 both funHddc and funWeightClust give good results. For the more difficult clustering problem in scenario 2, taking into account the relationship between $Y_i$ and $X_i$ seems to matter and only  funWeightClust succeeds to find the correct partition.
  %%%%%%%%%%%%%%%%%%%%%%%
%%%%%%%%%%%%%%%%%%%%%%%
%%%%%%%%%%%%%%%%%%%%%%%%%%%%%%%%%%%
%%%%%%%%%%%%%%%%%%%%%%%%%%%%%%%%%%%%%%%
\begin{table}[h!]
\caption{Mean (and standard deviation) of ARI for BIC best model on 100 simulations. Bold values indicate the highest value for each method.}
\label{table1}
\begin{tabular}{@{}llll@{}}
\toprule
Scenario&Method&$\epsilon$ &ARI \\
\midrule
1&FunHDDC&0.005& {\bf{0.8589641 (0.1401613)}}\\
1&FunHDDC&0.01& 0.8527159(0.1375641)\\
1&FunHDDC&0.2&0.7264915 (0.1236281)\\
1&2D-funHDDC&0.005&{\bf{0.2791434 (0.04373348)}}\\
1&2D-funHDDC&0.01&0.2791434(0.04373348)\\
1&2D-funHDDC&0.2&0.270681 (0.05636388)\\
1&funWeightClust&0.005& {\bf{0.9965396 (0.006011004)}}\\
1&funWeightClust&0.01& 0.9960756(0.006813709)\\
1&funWeightClust&0.2&0.1535521 (0.0371209)\\
\midrule
2&FunHDDC&0.005& 0.1309167 (0.04842073)\\
2&FunHDDC&0.01& {\bf{0.1309442(0.04844203)}}\\
2&FunHDDC&0.2&0.1220269 (0.04650234)\\
2&2D-funHDDC&0.005&0.1256394 (0.02688633)\\
2&2D-funHDDC&0.01&0.1256394(0.02688633)\\
2&2D-funHDDC&0.2&{\bf{0.1462503 (0.04981952)}}\\
2&funWeightClust&0.005& 0.8333807 (0.1946126)\\
2&funWeightClust&0.01& {\bf{0.8380116(0.1856003)}}\\
2&funWeightClust&0.2&0.06956302(0.0255526)\\
\botrule
\end{tabular}
\end{table}
%%%%%%%%%%%%%%%%%%%%%%%
%%%%%%%%%%%%%%%%%%%%%%% 

\subsection{Benchmark study- Adelaide electricity demand data}
%%%%%%%%%%%%%%%%%%%%%%%%%%%%%%%%%%%%%%%%
%%%%%%%%%%%%%%%%%%%%%%%%%%%%%%%%%%%%%%%
%%%%%%%%%%%%%%%%%%%%%%%%%%%%%%%%%%%%%%
The Adelaide data is available in the {\it fds} package in R and consist of  electricity demands, in Megawatt (MW) from Sunday to Saturday in Adelaide, Australia for 508 weeks, between July 6, 1976 and March 31, 2007  \citep{MagnanoBolandHyndman:2008}. The electricity demand is measured half-hourly, i.e. we have  48 time points in a day, and we have 508$\times $ 7 daily curves. These data were used to fit  concurrent functional regression models \citep{RamsaySilverman:2006} and non-linear function-on-function regression models using neural networks for the dependency of electricity demand on temperature \citep{RaoReimherr:2021}.  
\begin{figure}[h!]
\begin{center}
\includegraphics[width=12cm,height=8cm]{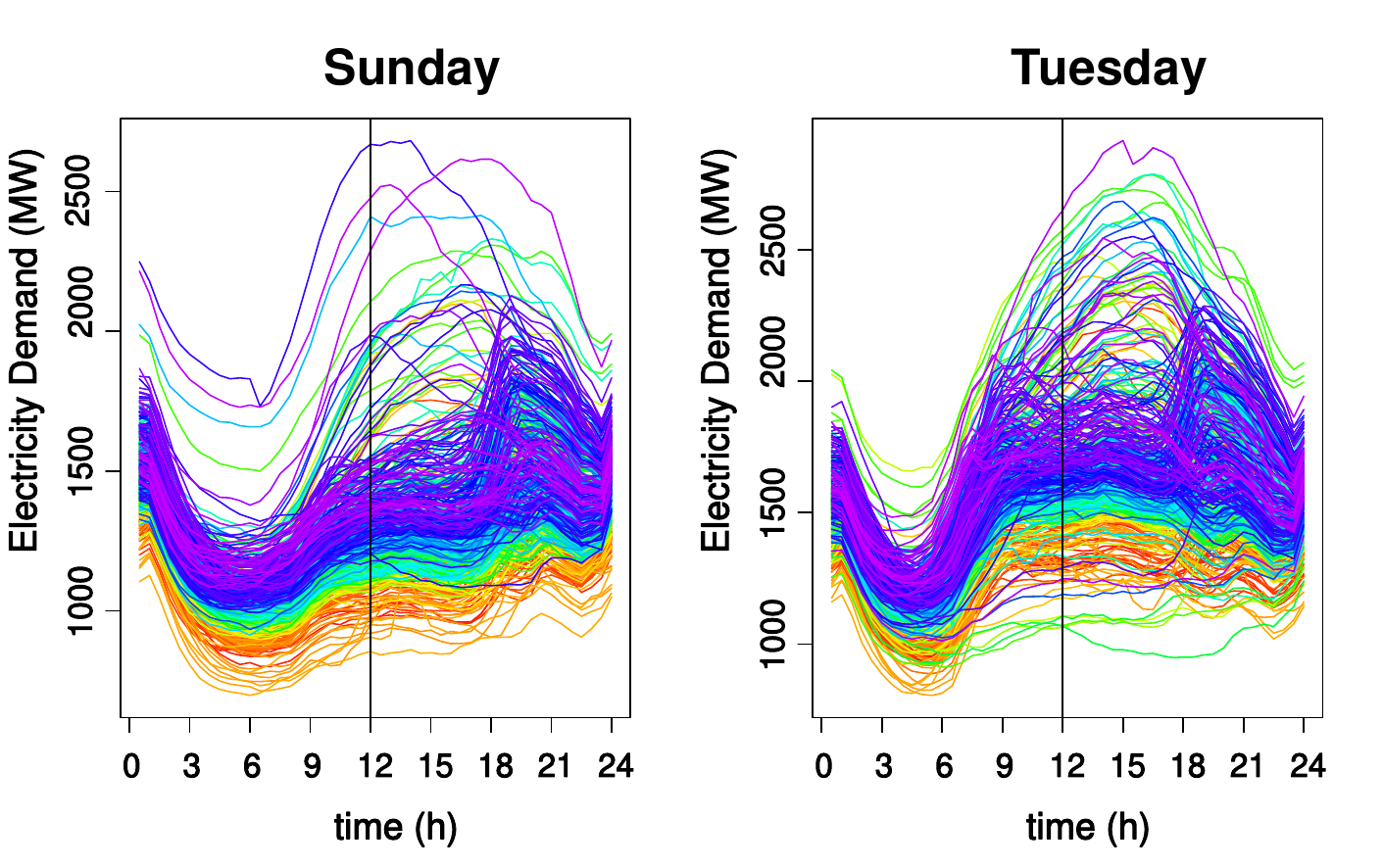}
\end{center}
  \caption{Electricity demand in Adelaide for Sundays and Tuesdays, between July 6, 1976 and March 31, 2007}
\label{fig3}
\end{figure}   

We restrict our analysis to Sundays and Tuesdays, so we have 1016 daily curves  (see Figure \ref{fig3}). We suppose that the electricity demand in the  morning can be used to predict the demand in the afternoon, so the covariate $X_i$ includes the first 24 points (from midnight to noon)  and  the last 24 points (from noon to midnight) are measurements corresponding to the response $Y_i$.  Electricity demand follows different dynamics on weekends (Sunday) compared to weekdays (Tuesday), so we apply funWeightClust to partition into these two groups. 
 
For comparison we also apply funHDDC to the combined curves and 2D-funHDDC, where, as before, by 2D-funHDDC we denote funHDDC applied to the two-dimensional curves $(X_i,Y_i)$. Curves are smoothed using  cubic B-splines with 6 basis elements. We run the algorithms for $K=2$ clusters using $\epsilon \in\{ 0.01, 0.1, 0.2\}$ in the Cattell test, and the best solution in terms of the highest BIC value for all sub-models is returned. The initialization is done using {{\it{kmeans}}} method with 20 repetitions, and the maximum number of iterations is 200.
 
We compare funWeightClust also with {\it{kmeans}} (from the R {\it stats} package), flexmix (from the R package {\it flexmix}), and Mclust (from the {\it mclust} R package), applied to the raw data and also applied as two-steps methods to the coefficients of the cubic B-spline basis with 6 basis elements. For flexmix we consider the data corresponding to the pairs $(X_i,Y_i)$, and for the other methods we work with the data corresponding to the combined curves defined on the whole interval $[0,24]$. 
%%%%%%%%%%%%%%%%%%%%%%%%%%%%%%
%%%%%%%%%%%%%%%%%%%%%%%
\begin{table}[h!]
\begin{center}
\begin{minipage}{\textwidth}
\caption{ARI for each method for the Adelaide data}
\label{table2}
\begin{tabular}{@{}llllllll@{}}
\toprule
Method&$\epsilon$&ARI&Method&$\epsilon$&ARI&Method&ARI\\
\midrule
funHDDC&0.01&0.48&2D-funHDDC&0.2&{\bf 0.53}&{\it{kmeans}}-two-steps method&{\bf 0.55}\\
funHDDC&0.1&{\bf 0.50}&funWeightClust&0.01&0.61&flexmix-raw data&-0.000\\
funHDDC&0.2&0.48 &funWeightClust&0.1&{\bf 0.94}&flexmix-two-steps method&-0.000\\
2D-funHDDC&0.01&0.34&funWeightClust&0.2&0.85&Mclust-raw data&0.002\\
2D-funHDDC&0.1&0.47&{\it{kmeans}}-raw data&-&0.48&Mclust-two-steps method&0.001\\
\botrule
\end{tabular}
\end{minipage}
\end{center}
\end{table}
%%%%%%%%%%%%%%%%%%%%%%%
%%%%%%%%%%%%%%%%%%%
 
\begin{figure}[h!]
\begin{center}
\includegraphics[width=12cm,height=8cm]{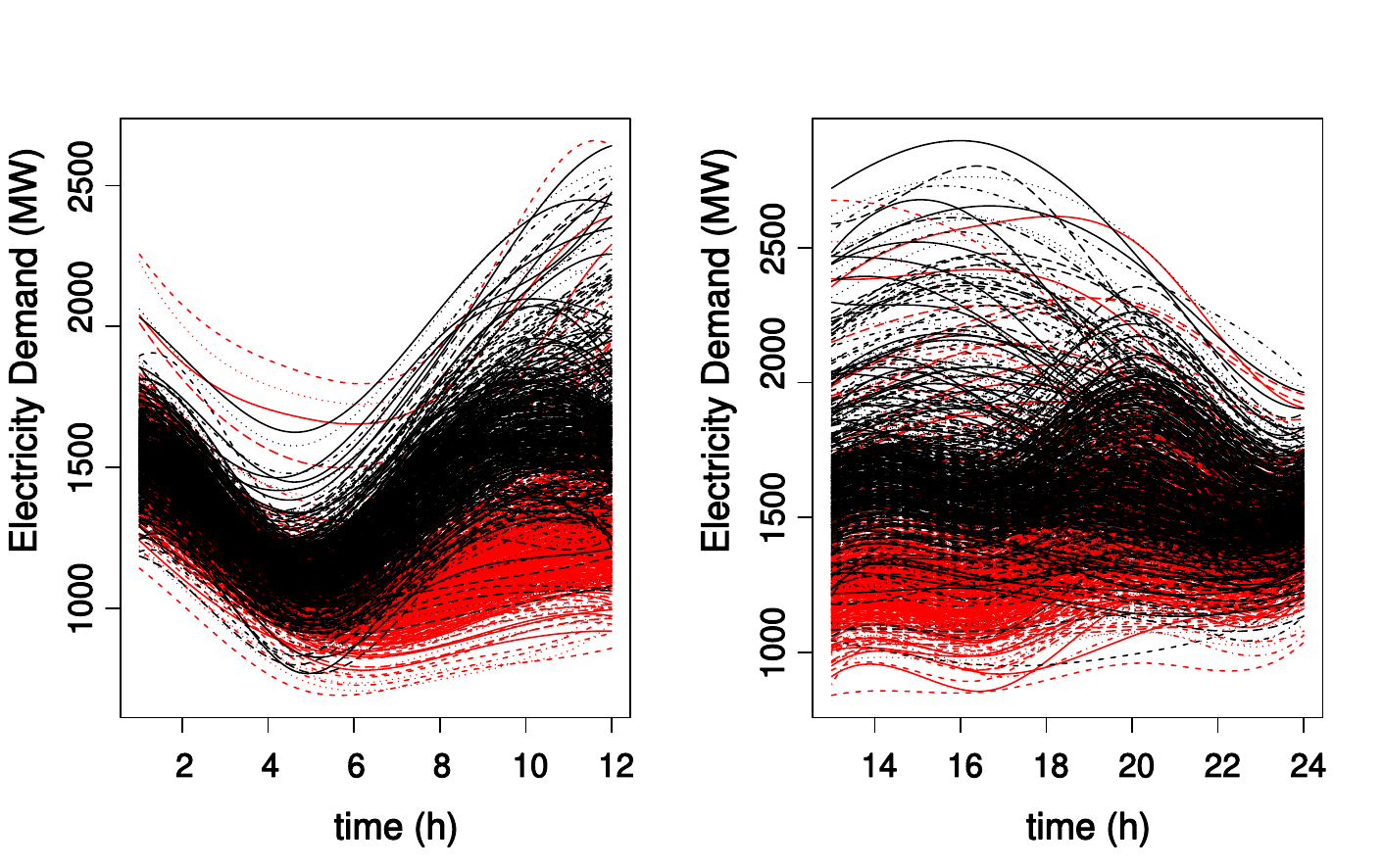}
\end{center}
  \caption{Clustering with funWeightClust  ($\epsilon=0.1$) of electricity demand in Adelaide for Sundays (red) and Tuesdays (black)}
\label{fig4}
\end{figure}

The results in Table \ref{table2} clearly show that funWeightClust outperforms the other methods, and for $\epsilon=0.1$ we obtain ARI=0.94, so a very accurate clustering. The clusters obtained with funWeightClust with $\epsilon=0.1$ are presented in Figure \ref{fig4}.  

From the results in Table \ref{table2} it seems that the difference between the electricity demand on Sundays and Tuesday is illustrated the best by the different relationship between the electricity demand in the mornings ($X_i$) and in the afternoons ($Y_i$). The only methods that consider the dependency between $Y_i$ and $X_i$ are flexmix and  funWeightClust, but flexmix does not account for correlated multivariate response variables and gives poor results for both the raw data and the data represented by the coefficients of the cubic B-splines. funHDDC and {\it{kmeans}} give the second best results, but these methods cluster the curves defined on the whole interval $[0,24]$. 
%%%%%%%%%%%%%%%%%%%%%%%%%%%
%%%%%%%%%%%%%%%%%%%%%%%%%%
%%%%%%%%%%%%%%%%%%%%%%%%%
\subsection{Real data example- Traffic Speeding}
%%%%%%%%%%%%%%%%%%%%
%%%%%%%%%%%%%%%%%%%%%
%%%%%%%%%%%%%%%%%%%%%%%%%
We use funWeightClust on speeding differential data extracted from speed surveys conducted by the City of Edmonton during the summer months of 2017 and 2018 available from an open data portal
(\url{https://data.edmonton.ca/stories/s/Speed-Surveys/kd7n-5iq3/}). We expect patterns to appear between the prediction of afternoon traffic (8:15 am – 11:45 pm) from the morning traffic rush (12 am – 8 am). Traffic data for car going 5 – 10 km/h under the speed limit (bin 2) and 0 – 5 km/h over the speed limit (bin 4) are considered. The $X_i$ part of the regression consists of the bin 2 and bin 4 morning traffic data, and the $Y_i$ component is the afternoon traffic, making this a multivariate traffic functional regression clustering problem. Data are composed of car counts recorded every 15 minutes. The observations are recorded for each $X_i$ bi-dimensional curve  at 36 time points (12 am – 8 am) and for each $Y_i$ bi-dimensional curve  at the remaining 60 time points (8:15 am – 11:45 pm).

A thousand records are sampled from the original data set for clustering with the funWeightClust method. We take the best result using the BIC criterion and compare the clustered curves by their locations in the city and speed limits. The objective of this study is to find patterns of traffic regressions related to travel preferences on roads. These results are the first step of an in-depth analysis of how morning and evening traffic affect each other and the trends that appear in this relationship.

Data are fit with a B-Spline basis using 6 basis functions for each component of the $X_i$  and $Y_i$ functional data. We use all models of the funWeightClust method, considering a range of clusters from 2 to 10. The best result was 4 clusters and a threshold for the Cattell test $\epsilon = 0.001$.

\begin{figure}[h!]
\begin{center}
\includegraphics[width=12cm,height=12cm]{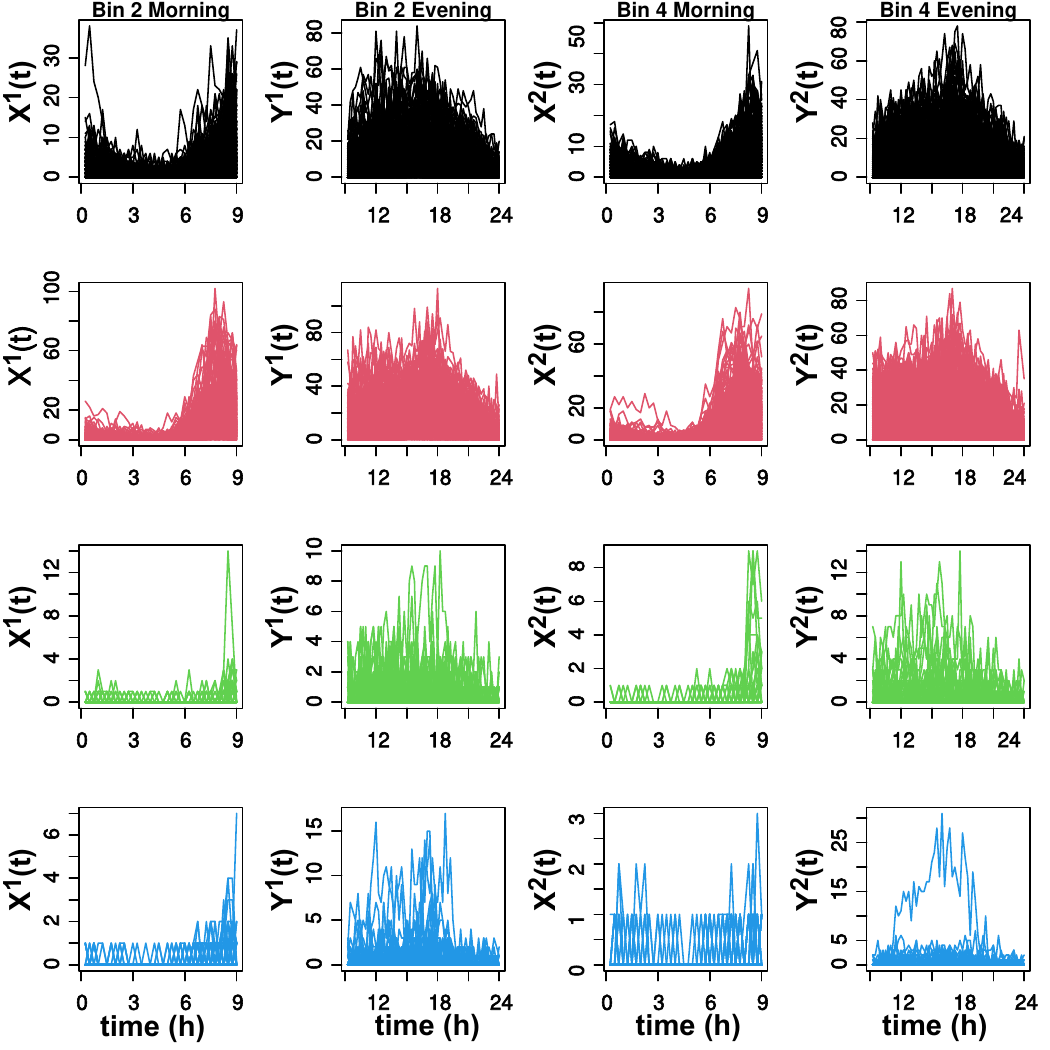}
\end{center}
  \caption{Clustering with funWeightClust of Edmonton's traffic data}
\label{fig5}
\end{figure}

The funWeightClust method clusters the data according to the different ways the traffic in the morning rush predicts the afternoon traffic. Within the four clusters, shown in Figure \ref{fig5}, we discern two large clusters (1 and 2) and two smaller clusters (3 and 4). Each cluster has distinct patterns of behavior of peaking only in the evening or peaking in the evening and morning. Clusters 3 and 4 have low amounts of travel, creating different patterns of behavior than clusters 1 and 2 in the response variable. From table \ref{table3} we see that while most of the data is recorded on roads with maximum speed limits of 50 and 60 km/h, the data for groups 3 and 4  are recorded mostly on roads with maximum speed limits of 50 and 30 km/h. Using latitude and longitude we identified that roads from clusters 3 and 4 are close to main roads networks. These roads may be neighborhood roads or side roads used as detours during rush hour. 
\begin{table}[h!]
\begin{center}
\begin{minipage}{\textwidth}
\caption{Maximum speed limit for the roads where the measurements were done for each cluster}
\label{table3}
\begin{tabular}{@{}lllll@{}}
\toprule
Maximum speed limit&cluster 1&cluster 2&cluster 3&cluster 4\\
\midrule
30&37&0&14&10\\
40&6&0&0&0\\
50&230&66&84&51\\
60&252&226&4&7\\
70&4&9&0&0\\
\botrule
\end{tabular}
\end{minipage}
\end{center}
\end{table}

%%%%%%%%%%%%%%%%%%%%%%%%%%%%%%%%%
%%%%%%%%%%%%%%%%%%%%%%%%%%%%%%%%
%%%%%%%%%%%%%%%%%%%%%%%%%%%%%
\section{Conclusions and future work}
\label{section5}  
We propose the method funWeightClust for clustering heterogeneous functional linear regression data. Based on a  cluster weighted model for functional data  and on a multivariate functional principal component analysis,  funWeightClust has the important advantage that  it can handle functional multivariate responses and predictors. We include the distribution of the coefficients of the covariates in the likelihood, and  we construct  an EM algorithm to estimate the parameters. Despite the complexity of the model, the maximization of the complete log likelihood can be done in closed form expressions. To add model flexibility, we consider several two-component parsimonious models by combining the parsimonious models used for funHDDC  \citep{ BouveyronJacques:2011}  with the Gaussian parsimonious clustering models family in  \cite{CeleuxGovaert:1995}.

We conduct experiments for simulated data and  for the Adelaide electricity demand data in the {\it  fds} R package, and compare the proposed method funWeightClust with funHDDC and two-steps methods based on clustering methods for multivariate data.  For simulated data that include linear regression dependencies between the variables, funWeightClust outperforms funHDDC. The differences between the Adelaide electricity demand during Sundays and Tuesday are captured very precisely by funWeightClust, with an ARI=0.94, much larger than the results obtained using the other methods. 

We also use funWeightClust to cluster speeding data in Edmonton, Alberta.  The proposed clusterwise functional regression model captures  the dependence of the afternoon traffic on the morning traffic. Using funWeightClust we were able to identify groups that show how the roads are utilized during the morning and afternoon rush hours. For cities like Edmonton, the analysis of traffic flow is crucial in transportation planning  and optimization of road design.

The proposed model is based on multivariate Gaussian distributions. Our approach can be extended  to non-Gaussian mixture models by considering multivariate
skewed distributions as in \cite{GallaugherTomarchioMcNicholasPunzo:2022}. 
Here we focus on unsupervised learning, but a similar methodology can be applied for functional classification and prediction (\citealp{ChiouYangChen:2016}, \citealp{Chiou:2012}).

%%%%%%%%%%%%%%%%%%%%%%%%%%%%%%%
%%%%%%%%%%%%%%%%%%%%%%%%%%%%%
%%%%%%%%%%%%%%%%%%%%%%%%%%
%%%%%%%%%%%%%%%%%%%%%%%%%%%%%%%%%%%%
%%%%%%%%%%%%%%%%%%%%%%%%%%%%%
\begin{appendices}
\section{Proof of Proposition \ref{proplik}}\label{secA1}
\begin{proof}
The complete-data likelihood can be written as the product of the conditional densities of the multivariate response $\mathbfit{c}_{Y,i}$ given the covariates ${\mathbfit{c}}_{X,i}$ and $\mathbfit{Z}_i=\mathbfit{z}_i$,  the conditional densities of ${\mathbfit{c}}_{X,i}$ given that $\mathbfit{Z}_i=\mathbfit{z}_i$, and the marginal densities of the  $\mathbfit{Z}_i$:
\begin{align*}
L_c(\boldsymbol{\theta})=\prod_{i=1}^n\prod_{k=1}^K\left\{\phi({\mathbfit{c}}_{Y,i};\boldsymbol{\mu}_{Y,k},\boldsymbol{\Sigma}_{Y,k})\phi({\mathbfit{c}}_{X,i};\boldsymbol{\mu}_{X,k},\boldsymbol{\Sigma}_{X,k})\pi_k\right\}^{z_{ik}},
\end{align*}
where $z_{ik}=1$ if $(\mathbfit{c}_{Y,i}, \mathbfit{c}_{X,i})$ belongs to the cluster $k$ and $z_{ik}=0$ otherwise.
Thus, the complete-data  log-likelihood can be written as
\begin{equation*}
l_c(\boldsymbol{\theta})=l_{1c}(\pi)+l_{2c}(\boldsymbol{\vartheta}_X)+l_{3c}(\boldsymbol{\vartheta}_Y)
\end{equation*}
where
\begin{align}
&l_{1c}(\pi)=\sum_{i=1}^n\sum_{k=1}^K z_{ik}\log(\pi_k)\nonumber\\
&l_{2c}(\boldsymbol{\vartheta}_X)=-\frac{1}{2}\sum_{i=1}^n\sum_{k=1}^K z_{ik}\biggl(R_X\log(2\pi)+\log\mid \boldsymbol{\Sigma}_{X,k}\mid +(\mathbfit{c}_{X,i}-\boldsymbol{\mu}_{X,k})^\top\boldsymbol{\Sigma}_{X,k}^{-1}(\mathbfit{c}_{X,i}-\boldsymbol{\mu}_{X,k})\biggl)\label{l3use}\\
&l_{3c}(\boldsymbol{\vartheta}_Y)=-\frac{1}{2}\sum_{i=1}^n\sum_{k=1}^K z_{ik}\biggl(R_Y\log(2\pi)+\log\mid \boldsymbol{\Sigma}_{Y,k}\mid +(\mathbfit{c}_{Y,i}-\boldsymbol{\mu}_{Y,k})^\top\boldsymbol{\Sigma}_{Y,k}^{-1}(\mathbfit{c}_{Y,i}-\boldsymbol{\mu}_{Y,k})\biggl)\nonumber\\
&=-\frac{1}{2}\sum_{i=1}^n\sum_{k=1}^K z_{ik}\biggl(R_Y\log(2\pi)+\log\mid \boldsymbol{\Sigma}_{Y,k}\mid +(\mathbfit{c}_{Y,i}-\boldsymbol{\Gamma}^k_{*}{\mathbfit{c}_{X,i}^{*}})^\top\boldsymbol{\Sigma}_{Y,k}^{-1}\nonumber\\
&(\mathbfit{c}_{Y,i}-\boldsymbol{\Gamma}^k_{*}{\mathbfit{c}_{X,i}^{*}})\biggl)\label{l2use}
\end{align}
%%%%%%%%%%%%%%%%%%%%%%%%%%
From \eqref{dis5} we have 
$$
\boldsymbol{\Sigma}_{X,k}^{-1}=\mathbfit{W}_X^{1/2}\mathbfit{Q}_k\mathbfit{D}_k^{-1}\mathbfit{Q}_k^\top \mathbfit{W}_X^{1/2},
$$
and
\begin{equation} 
\mid \boldsymbol{\Sigma}_{X,k}\mid =\mid \mathbfit{D}_k\mid \mid \mathbfit{W}_X\mid ^{-1}\mid \mid \mathbfit{Q}_k^\top\mathbfit{Q}_k\mid  =\mid \mathbfit{D}_k\mid \mid \mathbfit{W}_X\mid ^{-1} =\mid \mathbfit{W}_X\mid ^{-1}\prod_{l=1}^{d_k}a_{kl}\prod_{l=d_k+1}^{R_X} b_k.\label{dety}
\end{equation}
Moreover, since $(\mathbfit{c}_{X,i}-\boldsymbol{\mu}_{X,k})^\top\boldsymbol{\Sigma}_{X,k}^{-1}(\mathbfit{c}_{X,i}-\boldsymbol{\mu}_{X,k})$ is a scalar, we get
\begin{align}
&(\mathbfit{c}_{X,i}-\boldsymbol{\mu}_{X,k})^\top\boldsymbol{\Sigma}_{X,k}^{-1}(\mathbfit{c}_{X,i}-\boldsymbol{\mu}_{X,k})=\text{trace}((\mathbfit{c}_{X,i}-\boldsymbol{\mu}_{X,k})^\top\mathbfit{W}_X^{1/2}\mathbfit{Q}_k\mathbfit{D}_k^{-1}\mathbfit{Q}_k^\top \mathbfit{W}_X^{1/2}\nonumber\\
&(\mathbfit{c}_{X,i}-\boldsymbol{\mu}_{X,k})=\text{trace}\left(\left((\mathbfit{c}_{X,i}-\boldsymbol{\mu}_{X,k})^\top\mathbfit{W}_X^{1/2}\mathbfit{Q}_k\right)\left(\mathbfit{D}_k^{-1}\mathbfit{Q}_k^\top \mathbfit{W}_X^{1/2}(\mathbfit{c}_{X,i}-\boldsymbol{\mu}_{X,k})\right)\right)\nonumber\\
&=\text{trace}\left(\left(\mathbfit{D}_k^{-1}\mathbfit{Q}_k^\top \mathbfit{W}_X^{1/2}(\mathbfit{c}_{X,i}-\boldsymbol{\mu}_{X,k})\right)\left((\mathbfit{c}_{X,i}-\boldsymbol{\mu}_{X,k})^\top\mathbfit{W}_X^{1/2}\mathbfit{Q}_k\right)\right)\label{delty}
\end{align}
Replacing  in \eqref{l3use} we obtain
\begin{align*}
&l_{2c}(\boldsymbol{\vartheta}_X)=-\frac{nR_X\log(2\pi)}{2}+\frac{n}{2}\log(\mid \mathbfit{W}_X\mid)-\frac{1}{2}\sum_{k=1}^K n_k\sum_{l=1}^{d_k}\log(a_{kl})-\frac{1}{2}\sum_{k=1}^K n_k\sum_{l=d_k+1}^{R_X}\log(b_{k})\\
&-\frac{1}{2}\sum_{k=1}^K \text{trace} \biggl(\left(\mathbfit{D}_k^{-1}\mathbfit{Q}_k^\top \mathbfit{W}_X^{1/2}\right)\left(\sum_{i=1}^n z_{ik}(\mathbfit{c}_{X,i}-\boldsymbol{\mu}_{X,k})(\mathbfit{c}_{X,i}-\boldsymbol{\mu}_{X,k})^\top\right)\left(\mathbfit{W}_X^{1/2}\mathbfit{Q}_k\right)\biggl).
\end{align*}
We can rewrite $l_{2c}(\boldsymbol{\vartheta}_X)$ as 
\begin{align*}
&l_{2c}(\boldsymbol{\vartheta}_X)=-\frac{nR_X\log(2\pi)}{2}+\frac{n}{2}\log(\mid \mathbfit{W}_X\mid)-\frac{1}{2}\sum_{k=1}^K n_k\sum_{l=1}^{d_k}\log(a_{kl})\\
&-\frac{1}{2}\sum_{k=1}^K n_k\sum_{l=d_k+1}^{R_X}\log(b_{k})-\frac{1}{2}\sum_{k=1}^K  \text{trace} \biggl(\mathbfit{D}_k^{-1}\mathbfit{Q}_k^\top \mathbfit{W}_X^{1/2}\mathbfit{S}_{X,k}\mathbfit{W}_X^{1/2}\mathbfit{Q}_k\biggl)\\
&=-\frac{nR_X\log(2\pi)}{2}+\frac{n}{2}\log(\mid \mathbfit{W}_X\mid)-\frac{1}{2}\sum_{k=1}^K n_k\sum_{l=1}^{d_k}\log(a_{kl})-\frac{1}{2}\sum_{k=1}^K n_k\sum_{l=d_k+1}^{R_X}\log(b_{k})\\
&-\frac{1}{2}\sum_{k=1}^K   \biggl(\sum_{l=1}^{d_k}\frac{\mathbfit{q}_{kl}^\top \mathbfit{W}_X^{1/2}\mathbfit{S}_{X,k}\mathbfit{W}^{1/2}\mathbfit{q}_{kl}}{a_{kl}}+\sum_{l=d_k+1}^{R_X}\frac{\mathbfit{q}_{kl}^\top \mathbfit{W}_X^{1/2}\mathbfit{S}_{X,k}\mathbfit{W}_X^{1/2}\mathbfit{q}_{kl}}{b_{k}}\biggl),\\
\end{align*}
where $\mathbfit{q}_{kl}$ is the $l$th column of $\mathbfit{Q}_k$, $\mathbfit{S}_{X,k}$ is defined in \eqref{defS}, and $\boldsymbol{\vartheta}_X=\{\boldsymbol{\mu}_{X,k}, a_{kj}, b_k, \mathbfit{q}_{kj}\}$, $k=1,\ldots, k$, $j=1,\ldots, d_k$. 

Next, from \eqref{l2use} we have
\begin{align*}
&l_{3c}(\boldsymbol{\vartheta}_Y)=-\frac{nR_y\log(2\pi)}{2}-\frac{1}{2}\sum_{k=1}^K n_k\log(\log\mid \boldsymbol{\Sigma}_{Y,k}\mid )-\frac{1}{2}\sum_{i=1}^n\sum_{k=1}^K z_{ik}\biggl(\mathbfit{c}_{Y,i}^\top\boldsymbol{\Sigma}_{Y,k}^{-1}\mathbfit{c}_{Y,i}\nonumber\\
&-\mathbfit{c}_{Y,i}^\top\boldsymbol{\Sigma}_{Y,k}^{-1}\boldsymbol{\Gamma}^k_{*}{\mathbfit{c}_{X,i}^{*}}-({\mathbfit{c}_{X,i}^{*}})^\top\left(\boldsymbol{\Gamma}^k_{*}\right)^\top \boldsymbol{\Sigma}_{Y,k}^{-1}\mathbfit{c}_{Y,i}+({\mathbfit{c}_{X,i}^{*}})^\top\left(\boldsymbol{\Gamma}^k_{*}\right)^\top\boldsymbol{\Sigma}_{Y,k}^{-1}\boldsymbol{\Gamma}^k_{*}{\mathbfit{c}_{X,i}^{*}}\biggl)
\end{align*}
\end{proof}
%%%%%%%%%%%%%%%%%%%%%%%
%%%%%%%%%%%%%%%%%%%%%%%%
%%%%%%%%%%%%%%%%%%%%%%%%%%
\section{Proof of Proposition \ref{propEM}}
\label{secA2}
\begin{proof}
From \eqref{delty} we obtain
\begin{align}
&(\mathbfit{c}_{X,i}-\boldsymbol{\mu}_{X,k})^\top\boldsymbol{\Sigma}_{X,k}^{-1}(\mathbfit{c}_{X,i}-\boldsymbol{\mu}_{X,k})\\
&=\biggl(\sum_{l=1}^{d_k}\frac{\mathbfit{q}_{kl}^\top \mathbfit{W}_X^{1/2}(\mathbfit{c}_{X,i}-\boldsymbol{\mu}_{X,k})(\mathbfit{c}_{X,i}-\boldsymbol{\mu}_{X,k})^\top \mathbfit{W}_X^{1/2}\mathbfit{q}_{kl}}{a_{kl}}\nonumber\\
&+\sum_{l=d_k+1}^{R}\frac{\mathbfit{q}_{kl}^\top \mathbfit{W}_X^{1/2}(\mathbfit{c}_{X,i}-\boldsymbol{\mu}_{X,k})(\mathbfit{c}_{X,i}-\boldsymbol{\mu}_{X,k})^\top\mathbfit{W}_X^{1/2}\mathbfit{q}_{kl}}{b_{k}}\biggl)\nonumber\\
&=\delta(\mathbfit{c}_{X,i};\boldsymbol{\mu}_{X,k},\mathbfit{Q}_k,a,b,d_k).
\end{align}
Replacing in \eqref{dens} and using also \eqref{densN}, \eqref{disty}, and \eqref{dety} we obtain
\begin{align*}
&p_k(\mathbfit{c}_{Y,i},\mathbfit{c}_{X,i} \mid \boldsymbol{\theta}_k)=\phi({\mathbfit{c}}_{X,i};\boldsymbol{\mu}_{X,k},\boldsymbol{\Sigma}_{X,k})\phi({\mathbfit{c}}_{Y,i};\boldsymbol{\mu}_{Y,k},\boldsymbol{\Sigma}_{Y,k})\\
&=(2\pi)^{-(R_X+R_Y)/2}\mid \boldsymbol{\Sigma}_{X,k}\mid ^{-1/2}\mid\boldsymbol{\Sigma}_{Y,k}\mid ^{-1/2}\exp\biggl(-\frac{1}{2}({\mathbfit{c}}_{X,i}-\boldsymbol{\mu}_{X,k})^\top\boldsymbol{\Sigma}_{X,k}^{-1}\\
&({\mathbfit{c}}_{X,i}-\boldsymbol{\mu}_{X,k})-\frac{1}{2}({\mathbfit{c}}_{Y,i}-\boldsymbol{\mu}_{Y,k})^\top\boldsymbol{\Sigma}_{Y,k}^{-1}({\mathbfit{c}}_{Y,i}-\boldsymbol{\mu}_{Y,k})\biggl)\\
%%%%%%%%%%%%%%%%%%%%%
&=(2\pi)^{-(R_X+R_Y)/2}\left(\prod_{j=1}^{d_k} a_{kj}\prod_{j=d_k+1}^{R_X}b_k\right)^{-1/2}\mid \mathbfit{W}_X\mid ^{1/2}\mid\boldsymbol{\Sigma}_{Y,k}\mid ^{-1/2}\\
&\exp\biggl(-\frac{1}{2}\delta(\mathbfit{c}_{X,i};\boldsymbol{\mu}_{X,k},\mathbfit{Q}_k,a,b,d_k)-\frac{1}{2}({\mathbfit{c}}_{Y,i}-\boldsymbol{\Gamma}^k_{*}{\mathbfit{c}_{X,i}^{*}} )^\top\boldsymbol{\Sigma}_{Y,k}^{-1}\\
&({\mathbfit{c}}_{Y,i}-\boldsymbol{\Gamma}^k_{*}{\mathbfit{c}_{X,i}^{*}} )\biggl)\\
%%%%%%%%%%%%%%%%%%%%%%%%%
&=(2\pi)^{-(R_X+R_Y)/2}\mid \mathbfit{W}_X\mid ^{1/2}\exp\biggl(-\frac{1}{2}\biggl(\sum_{j=1}^{d_k} \log (a_{kj})+(R_X-d_k)\log (b_k)\\
&+\log(\mid\boldsymbol{\Sigma}_{Y,k}\mid)+\delta(\mathbfit{c}_{X,i};\boldsymbol{\mu}_{X,k},\mathbfit{Q}_k,a,b,d_k)+({\mathbfit{c}}_{Y,i}-\boldsymbol{\Gamma}^k_{*}{\mathbfit{c}_{X,i}^{*}} )^\top\boldsymbol{\Sigma}_{Y,k}^{-1}\\
&({\mathbfit{c}}_{Y,i}-\boldsymbol{\Gamma}^k_{*}{\mathbfit{c}_{X,i}^{*}} )\biggl)\biggl)\\
&=(2\pi)^{-(R_X+R_Y)/2}\mid \mathbfit{W}_X\mid ^{1/2}\pi_k^{-1}\exp\left(-\frac{1}{2}H_k(\mathbfit{c}_{Y,i},\mathbfit{c}_{X,i}\mid \boldsymbol{\theta}_k)\right),
\end{align*}
where $H_k(\mathbfit{c}_{Y,i},\mathbfit{c}_{X,i}\mid \boldsymbol{\theta}_k)$ is defined in \eqref{Hk}.
\end{proof}
%%%%%%%%%%%%%%%%%%%%%%%%%%%%%%%%%%%%%%%%%%%%%%%
%%%%%%%%%%%%%%%%%%%%%%%%%%%%%%%%
%%%%%%%%%%%%%%%%%%%%%%%%%%%%%%%%%
%%%%%%%%%%%%%%%%%%%%%%%%%%%%%%%%%%%%%
\section{Proof of Proposition \ref{propM}}
\label{secA3}
\begin{proof}
Using \eqref{lik2}-\eqref{l2c2} we have that $Q(\boldsymbol{\theta}\mid \boldsymbol{\theta}^{(m-1)})$ is given by
\begin{equation*}
Q(\boldsymbol{\theta}\mid \boldsymbol{\theta}^{(m-1)})=Q_1(\pi\mid \boldsymbol{\theta}^{(m-1)})+Q_2(\boldsymbol{\vartheta}_X\mid \boldsymbol{\theta}^{(m-1)})+Q_3(\boldsymbol{\vartheta}_X\mid \boldsymbol{\theta}^{(m-1)}),
\end{equation*}
were 
\begin{align*}
&Q_1(\pi\mid \boldsymbol{\theta}^{(m-1)})=\sum_{i=1}^n\sum_{k=1}^K t_{ik}^{(m)}\log(\pi_k)\\
&Q_2(\boldsymbol{\vartheta}_X\mid \boldsymbol{\theta}^{(m-1)})=-\frac{nR_X\log(2\pi)}{2}+\frac{n}{2}\log(\mid \mathbfit{W}_X\mid)\\
&-\frac{1}{2}\sum_{k=1}^K n_k^{(m)}\sum_{l=1}^{d_k}\log(a_{kl})-\frac{1}{2}\sum_{k=1}^K n_k^{(m)}\sum_{l=d_k+1}^{R_X}\log(b_{k})\\
&-\frac{1}{2}\sum_{k=1}^K n_k^{(m)}  \biggl(\sum_{l=1}^{d_k}\frac{\mathbfit{q}_{kl}^\top \mathbfit{W}_X^{1/2}\mathbfit{S}_{X,k}^{(m)}\mathbfit{W}_X^{1/2}\mathbfit{q}_{kl}}{a_{kl}}+\sum_{l=d_k+1}^{R_X}\frac{\mathbfit{q}_{kl}^\top \mathbfit{W}_X^{1/2}\mathbfit{S}_{X,k}^{(m)}\mathbfit{W}_X^{1/2}\mathbfit{q}_{kl}}{b_{k}}\biggl),\\
&Q_3(\boldsymbol{\vartheta}_Y\mid \boldsymbol{\theta}^{(m-1)})=-\frac{nR_y\log(2\pi)}{2}-\frac{1}{2}\sum_{k=1}^K n_{k}^{(m)}\log(\mid \boldsymbol{\Sigma}_{Y,k}\mid) \\
&-\frac{1}{2}\sum_{i=1}^n\sum_{k=1}^K t_{ik}^{(m)}(\mathbfit{c}_{Y,i}-\boldsymbol{\Gamma}^k_{*}{\mathbfit{c}_{X,i}^{*}})^\top\boldsymbol{\Sigma}_{Y,k}^{-1}
(\mathbfit{c}_{Y,i}-\boldsymbol{\Gamma}^k_{*}{\mathbfit{c}_{X,i}^{*}})\biggl)
\end{align*}
where $\mathbfit{S}_{X,k}^{(m)}$ is defined in \eqref{sig}.

For the estimation of $\pi_k$, $k=1,\ldots, K$ we introduce the Lagrange multiplier $\lambda$ and we maximize 
$Q_1=Q_1(\pi\mid \boldsymbol{\theta}^{(m-1)})-\lambda(\sum_{k=1}^K \pi_k-1)$. We get \eqref{tpi} solving the system
\begin{equation*}
\frac{\partial Q_1}{\partial \pi_k}=\sum_{i=1}^n\frac{ t_{ik}^{(m)}}{\pi_k}-\lambda=0, k=1,\ldots, K\quad \frac{\partial Q_1}{\partial \lambda}=\sum_{k=1}^K \pi_k-1=0.
\end{equation*}

To get an update for $\boldsymbol{\boldsymbol{\mu}}_{X,k}^{(m)}$ we calculate $Q_2(\boldsymbol{\vartheta}_X\mid \boldsymbol{\theta}^{(m-1)})$ starting from the formula \eqref{l3use} and we obtain
\begin{align*}
Q_2(\boldsymbol{\vartheta}_X\mid \boldsymbol{\theta}^{(m-1)})&=-\frac{n}{2}R_X\log(2\pi)-\frac{1}{2}\sum_{k=1}^K n_{k}^{(m)}\log\mid \boldsymbol{\Sigma}_{X,k}\mid \nonumber\\
&-\frac{1}{2}\sum_{i=1}^n\sum_{k=1}^K t_{ik}^{(m)}(\mathbfit{c}_{X,i}-\boldsymbol{\boldsymbol{\mu}}_{X,k})^\top\boldsymbol{\Sigma}_{X,k}^{-1}(\mathbfit{c}_{X,i}-\boldsymbol{\boldsymbol{\mu}}_{X,k}).
\end{align*}
The gradient of $Q_2$ with respect to  ${\boldsymbol{\mu}}_{X,k}$ is
\begin{align*}
\nabla_{{\boldsymbol{\mu}}_{X,k}} Q_2(\boldsymbol{\vartheta}_X\mid \boldsymbol{\theta}^{(m-1)})&=-\sum_{i=1}^n t_{ik}^{(m)}\boldsymbol{\Sigma}_{X,k}^{-1}(\mathbfit{c}_{X,i}-\boldsymbol{\mu}_{X,k})\nonumber\\
&=\boldsymbol{\Sigma}_{X,k}^{-1}\left(-\sum_{i=1}^n t_{ik}^{(m)}\mathbfit{c}_{X,i}+{\boldsymbol{\mu}}_{X,k}\sum_{i=1}^n t_{ik}^{(m)}\right).
\end{align*}
Thus, we can easily get \eqref{tmu} solving $\nabla_{{\boldsymbol{\mu}}_{X,k}} Q_2(\boldsymbol{\vartheta}_X\mid \boldsymbol{\theta}^{(m-1)})={\bf 0}$.

To estimate $\mathbfit{Q}_k$ we have to maximize $Q_2(\boldsymbol{\vartheta}_X\mid \boldsymbol{\theta}^{(m-1)})$ with respect to $\mathbfit{q}_{kl}$ under the constraint   $\mathbfit{q}_{kl}^\top\mathbfit{q}_{kl}=1$. This is equivalent with minimizing $-2Q_2(\boldsymbol{\vartheta}_X\mid \boldsymbol{\theta}^{(m-1)})$ with respect to $\mathbfit{q}_{kl}$ under this constraint, so  we consider the  function $Q_{2c}=-2Q_2(\boldsymbol{\vartheta}_X\mid \boldsymbol{\theta}^{(m-1)})-\sum_{l=1}^{R_X} \omega_{kl}(\mathbfit{q}_{kl}^\top\mathbfit{q}_{kl}-1)$, where $\omega_{kl}$ are Lagrange multipliers.
The gradient of $Q_{2c}$ with respect to $\mathbfit{q}_{kl}$ is
\begin{align*}
\nabla_{\mathbfit{q}_{kl}}Q_{2c}&= 2n_k^{(m)}\frac{ \mathbfit{W}_X^{1/2}\mathbfit{S}_{X,k}^{(m)}\mathbfit{W}_X^{1/2}\mathbfit{q}_{kl}}{\Sigma_{kl}}-2\omega_{kl}\mathbfit{q}_{kl},\\
&\Sigma_{kl}=\begin{cases}&a_{kl}\text{ if } l=1,\ldots, d_k\\
&b_k\text{ if } l=d_k+1,\ldots, R_X.
\end{cases}
\end{align*}
From $\nabla_{\mathbfit{q}_{kl}}Q_{2c}=0$ we get $ \mathbfit{W}_X^{1/2}\mathbfit{S}_{X,k}^{(m)}\mathbfit{W}_X^{1/2}\mathbfit{q}_{kl}=\frac{\omega_{kl}\Sigma_{kl}}{n_k^{(m)}}\mathbfit{q}_{kl}$, so $\mathbfit{q}_{kl}$ is an eigenfunction of $\mathbfit{W}_X^{1/2}\mathbfit{S}_{X,k}^{(m)}\mathbfit{W}_X^{1/2}$ and the associated eigenvalue is $\lambda_{kl}^{(m)}=\frac{\omega_{kl}\Sigma_{kl}}{n_k^{(m)}}$. Notice that we also have $\mathbfit{q}_{kl}^\top\mathbfit{q}_{kj}=0$ if $l\ne j$, and  $\lambda_{kl}^{(m)}=\mathbfit{q}_{kl}^\top\mathbfit{W}_X^{1/2}\mathbfit{S}_{X,k}^{(m)}\mathbfit{W}_X^{1/2}\mathbfit{q}_{kl}$ so we can write
\begin{align*}
&-2Q_2(\boldsymbol{\vartheta}_X\mid \boldsymbol{\theta}^{(m-1)})=nR_X\log(2\pi)-n\log(\mid \mathbfit{W}_X\mid)+\sum_{k=1}^K n_k^{(m)}\biggl(\sum_{l=1}^{d_k}\log(a_{kl})\\
&+\sum_{l=d_k+1}^{R_X}\log(b_{k})\biggl)+\sum_{k=1}^K n_k^{(m)}  \biggl(\sum_{l=1}^{d_k}\frac{\lambda_{kl}^{(m)}}{a_{kl}}+\sum_{l=d_k+1}^{R_X}\frac{\lambda_{kl}^{(m)}}{b_{k}}\biggl)\\
&=nR_X\log(2\pi)-n\log(\mid \mathbfit{W}_X\mid)+\sum_{k=1}^K n_k^{(m)}\biggl(\sum_{l=1}^{d_k}\log(a_{kl})+\sum_{l=d_k+1}^{R_X}\log(b_{k})\biggl)\\
&+\sum_{k=1}^K n_k^{(m)}  \biggl(\sum_{l=1}^{d_k}\lambda_{kl}^{(m)}\left(\frac{1}{a_{kl}}-\frac{1}{b_k}\right)+\frac{1}{b_k} \text{trace}(\mathbfit{W}_X^{1/2}\mathbfit{S}_{X,k}^{(m)}\mathbfit{W}_X^{1/2})\biggl).
\end{align*}
Here we have also used 
\begin{equation}
\text{trace}(\mathbfit{W}_X^{1/2}\mathbfit{S}_{X,k}^{(m)}\mathbfit{W}_X^{1/2})=\sum_{l=1}^{R_X} \lambda_{kl}^{(m)}=\sum_{l=1}^{d_k} \lambda_{kl}^{(m)}+\sum_{l=d_k+1}^{R_X} \lambda_{kl}^{(m)}.\label{trac}
\end{equation}
Since for any $l=1,\ldots, d_k$ we have $a_{kl}\ge b_k$, we get $\frac{1}{a_{kl}}-\frac{1}{b_k}\le 0$, so $\sum_{l=1}^{d_k}\lambda_{kl}^{(m)}\left(\frac{1}{a_{kl}}-\frac{1}{b_k}\right)$ is a decreasing function of $\lambda_{kl}$. Thus,  we estimate $\mathbfit{q}_{kl}$ by the eigenfunction associated with the $l$th highest eigenvalue of $\mathbfit{W}_X^{1/2}\mathbfit{S}_{X,k}^{(m)}\mathbfit{W}_X^{1/2}$.

To update $a_{kl}$ we solve
\begin{align*}
\frac{\partial Q_2(\boldsymbol{\vartheta}_X\mid \boldsymbol{\theta}^{(m-1)})}{\partial a_{kl}}=-\frac{n_k^{(m)}}{2a_{kl}}+\frac{n_k^{(m)}\mathbfit{q}_{kl}^\top \mathbfit{W}^{1/2}\mathbfit{S}_{X,k}^{(m)}\mathbfit{W}^{1/2}\mathbfit{q}_{kl}}{2a_{kl}^2}=0,
\end{align*}
and we get $a_{kl}^{(m)}=\mathbfit{q}_{kl}^\top \mathbfit{W}_X^{1/2}\mathbfit{S}_{X,k}^{(m)}\mathbfit{W}_X^{1/2}\mathbfit{q}_{kl}=\lambda_{kl}^{(m)}$, the $l$th highest eigenvalue of $\mathbfit{W}_X^{1/2}\mathbfit{S}_{X,k}^{(m)}\mathbfit{W}_X^{1/2}$.

From 
\begin{align*}
\frac{\partial Q_2(\boldsymbol{\vartheta}_X\mid \boldsymbol{\theta}^{(m-1)})}{\partial b_{k}}=-\frac{n_k^{(m)}}{2}\sum_{l=d_k+1}^{R_X}\frac{1}{b_{k}}+\frac{n_k^{(m)}}{2}\sum_{l=d_k+1}^{R_X}\frac{\mathbfit{q}_{kl}^\top \mathbfit{W}_X^{1/2}\mathbfit{S}_{X,k}^{(m)}\mathbfit{W}_X^{1/2}\mathbfit{q}_{kl}}{b_{k}^2}=0,
\end{align*}
we obtain
$$
b_{k}^{(m)}=\frac{1}{R_X-d_k}\sum_{l=d_k+1}^{R_X}\mathbfit{q}_{kl}^\top \mathbfit{W}_X^{1/2}\mathbfit{S}_{X,k}^{(m)}\mathbfit{W}_X^{1/2}\mathbfit{q}_{kl}=\frac{1}{R_X-d_k}\sum_{l=d_k+1}^{R_X}\lambda_{kl}^{(m)}
$$
Thus,  using \eqref{trac} we get
$$
b_{k}^{(m)}=\frac{1}{R_X-d_k}\left(\text{trace}(\mathbfit{W}_X^{1/2}\mathbfit{S}_{X,k}^{(m)}\mathbfit{W}_X^{1/2})-
\sum_{l=1}^{d_k} \lambda_{kl}^{(m)}\right).
$$ 

To estimate the regression coefficient we use the properties of trace and transpose and $\boldsymbol{\Sigma}_{Y,k}^{\top}=\boldsymbol{\Sigma}_{Y,k}$  and we get
\begin{align*}
&Q_3(\boldsymbol{\vartheta}_Y\mid \boldsymbol{\theta}^{(m-1)})=-\frac{nR_y\log(2\pi)}{2}-\frac{1}{2}\sum_{k=1}^K n_{k}^{(m)}\log(\mid \boldsymbol{\Sigma}_{Y,k}\mid) \\
&-\frac{1}{2}\sum_{i=1}^n\sum_{k=1}^K t_{ik}^{(m)}\biggl(\mathbfit{c}_{Y,i}^\top\boldsymbol{\Sigma}_{Y,k}^{-1}\mathbfit{c}_{Y,i}-\text{trace}\left(\mathbfit{c}_{Y,i}^\top\boldsymbol{\Sigma}_{Y,k}^{-1}\boldsymbol{\Gamma}^k_{*}{\mathbfit{c}_{X,i}^{*}}\right)\\
&-\text{trace}\left(({\mathbfit{c}_{X,i}^{*}})^\top\left(\boldsymbol{\Gamma}^k_{*}\right)^\top \boldsymbol{\Sigma}_{Y,k}^{-1}\mathbfit{c}_{Y,i}\right)+\text{trace}\left(({\mathbfit{c}_{X,i}^{*}})^\top\left(\boldsymbol{\Gamma}^k_{*}\right)^\top\boldsymbol{\Sigma}_{Y,k}^{-1}\boldsymbol{\Gamma}^k_{*}{\mathbfit{c}_{X,i}^{*}}\right)\biggl)\\
%%%%%%%%%%%%%%%%%%%%%%%%%%%%%%%%
&=-\frac{nR_y\log(2\pi)}{2}-\frac{1}{2}\sum_{k=1}^K n_{k}^{(m)}\log(\mid \boldsymbol{\Sigma}_{Y,k}\mid)
-\frac{1}{2}\sum_{i=1}^n\sum_{k=1}^K t_{ik}^{(m)}\biggl(\mathbfit{c}_{Y,i}^\top\boldsymbol{\Sigma}_{Y,k}^{-1}\mathbfit{c}_{Y,i}\\
&-\text{trace}\left(\boldsymbol{\Gamma}^k_{*}{\mathbfit{c}_{X,i}^{*}}\mathbfit{c}_{Y,i}^\top\boldsymbol{\Sigma}_{Y,k}^{-1}\right)-\text{trace}\left(\boldsymbol{\Sigma}_{Y,k}^{-1}\mathbfit{c}_{Y,i}({\mathbfit{c}_{X,i}^{*}})^\top\left(\boldsymbol{\Gamma}^k_{*}\right)^\top \right)\\
&+\text{trace}\left(\boldsymbol{\Gamma}^k_{*}{\mathbfit{c}_{X,i}^{*}}({\mathbfit{c}_{X,i}^{*}})^\top\left(\boldsymbol{\Gamma}^k_{*}\right)^\top\boldsymbol{\Sigma}_{Y,k}^{-1}\right)\biggl)\\
%%%%%%%%%%%%%%%%%%%%%%%%%%%%%
&=-\frac{nR_y\log(2\pi)}{2}-\frac{1}{2}\sum_{k=1}^K n_{k}^{(m)}\log(\mid \boldsymbol{\Sigma}_{Y,k}\mid)
-\frac{1}{2}\sum_{i=1}^n\sum_{k=1}^K t_{ik}^{(m)}\biggl(\mathbfit{c}_{Y,i}^\top\boldsymbol{\Sigma}_{Y,k}^{-1}\mathbfit{c}_{Y,i}\\
&-2\text{trace}\left(\boldsymbol{\Gamma}^k_{*}{\mathbfit{c}_{X,i}^{*}}\mathbfit{c}_{Y,i}^\top\boldsymbol{\Sigma}_{Y,k}^{-1}\right)+\text{trace}\left(\boldsymbol{\Gamma}^k_{*}{\mathbfit{c}_{X,i}^{*}}({\mathbfit{c}_{X,i}^{*}})^\top\left(\boldsymbol{\Gamma}^k_{*}\right)^\top\boldsymbol{\Sigma}_{Y,k}^{-1}\right)\biggl)
\end{align*}
To update $\boldsymbol{\Gamma}^k_{*}$ we solve
\begin{align}
&\frac{\partial Q_3(\boldsymbol{\vartheta}_Y\mid \boldsymbol{\theta}^{(m-1)})}{\partial \boldsymbol{\Gamma}^k_{*}}=\mathbf{0},\\
&-\frac{1}{2}\sum_{i=1}^n t_{ik}^{(m)}\left(-2\boldsymbol{\Sigma}_{Y,k}^{-1}\mathbfit{c}_{Y,i}({\mathbfit{c}_{X,i}^{*}})^\top+2\boldsymbol{\Sigma}_{Y,k}^{-1}\boldsymbol{\Gamma}^k_{*}{\mathbfit{c}_{X,i}^{*}}({\mathbfit{c}_{X,i}^{*}})^\top\right)=\mathbf{0}.
\end{align}
Thus for any $k=1,\ldots, K$ we obtain
\begin{equation*}
(\boldsymbol{\Gamma}^k_{*})^{(m)}=\left(\sum_{i=1}^n t_{ik}^{(m)}\mathbfit{c}_{Y,i}({\mathbfit{c}_{X,i}^{*}})^\top\right)\left(\sum_{i=1}^n t_{ik}^{(m)}{\mathbfit{c}_{X,i}^{*}}({\mathbfit{c}_{X,i}^{*}})^\top\right)^{-1}.
\end{equation*}
Notice that using again properties of trace and transpose we have
\begin{align*}
&Q_3(\boldsymbol{\vartheta}_Y\mid \boldsymbol{\theta}^{(m-1)})=-\frac{nR_y\log(2\pi)}{2}+\frac{1}{2}\sum_{k=1}^K n_{k}^{(m)}\log(\mid \boldsymbol{\Sigma}_{Y,k}^{-1}\mid) \\
&-\frac{1}{2}\sum_{i=1}^n\sum_{k=1}^K t_{ik}^{(m)}{\text{trace}}\left(({\mathbfit{c}}_{Y,i}-\boldsymbol{\Gamma}^k_{*}{\mathbfit{c}_{X,i}^{*}} )^\top\boldsymbol{\Sigma}_{Y,k}^{-1}({\mathbfit{c}}_{Y,i}-\boldsymbol{\Gamma}^k_{*}{\mathbfit{c}_{X,i}^{*}} )\right)\\
&=-\frac{nR_y\log(2\pi)}{2}+\frac{1}{2}\sum_{k=1}^K n_{k}^{(m)}\log(\mid \boldsymbol{\Sigma}_{Y,k}^{-1}\mid)\\
 &-\frac{1}{2}\sum_{i=1}^n\sum_{k=1}^K t_{ik}^{(m)}{\text{trace}}\left(\boldsymbol{\Sigma}_{Y,k}^{-1}({\mathbfit{c}}_{Y,i}-\boldsymbol{\Gamma}^k_{*}{\mathbfit{c}_{X,i}^{*}} )({\mathbfit{c}}_{Y,i}-\boldsymbol{\Gamma}^k_{*}{\mathbfit{c}_{X,i}^{*}} )^\top\right)
\end{align*}
Taking the derivative we obtain
\begin{align*}
&\frac{\partial Q_3(\boldsymbol{\vartheta}_Y\mid \boldsymbol{\theta}^{(m-1)})}{\partial \boldsymbol{\Sigma}_{Y,k}^{-1}}=\mathbf{0},\\
&\frac{1}{2}n_{k}^{(m)}\left(\boldsymbol{\Sigma}_{Y,k}^{-1}\right)^{-1}- \frac{1}{2}\sum_{i=1}^n t_{ik}^{(m)}({\mathbfit{c}}_{Y,i}-\boldsymbol{\Gamma}^k_{*}{\mathbfit{c}_{X,i}^{*}} )({\mathbfit{c}}_{Y,i}-\boldsymbol{\Gamma}^k_{*}{\mathbfit{c}_{X,i}^{*}} )^\top=\mathbf{0}.
\end{align*}
Thus
\begin{equation*}
\boldsymbol{\Sigma}_{Y,k}^{(m)}=\frac{\sum_{i=1}^n t_{ik}^{(m)}({\mathbfit{c}}_{Y,i}-(\boldsymbol{\Gamma}^k_{*})^{(m)}{\mathbfit{c}_{X,i}^{*}} )({\mathbfit{c}}_{Y,i}-(\boldsymbol{\Gamma}^k_{*})^{(m)}{\mathbfit{c}_{X,i}^{*}} )^\top}{n_{k}^{(m)}}
\end{equation*}
\end{proof}
%%%%%%%%%%%%%%%%%%%%%%%%%%%%%%%%
\section{Regression coefficients for simulations in Section \ref{simu}}
\label{apsimu}
$\boldsymbol{\Gamma}_0^1=(40.55785, -136.39392, 364.48536, -1009.44904, 137.08185, -53.19644)^\top$, $\boldsymbol{\Gamma}_0^2=(10.55038, -94.09524, 386.14923, -886.60180, 105.71551, 250.83997)^\top$,
\begin{align*}
&\boldsymbol{\Gamma}^1=\begin{pmatrix}
0.1604788& -0.006105744& -0.34002176 & 0.8045786& -1.809933 & 3.1085371 \\
0.0216076&  0.360282816&  0.02519028 & 0.0829196& -1.394305&  2.9290406\\
-1.5683676 & 1.279807057 &-2.81184706&  4.6599002 &-6.812895 & 7.9705907\\
3.2542660& -0.769508698 & 1.18667337 &-2.3600145&  2.725763 &-1.9297317\\
-2.1495590&  2.106397045& -0.45220970 &-0.5759578&  1.517537& -0.9133536\\
1.8634967& -1.706644614&  1.91048360& -1.6978476 & 1.125397& -0.1015894
\end{pmatrix},\\
&\boldsymbol{\Gamma}^2=\begin{pmatrix}
-0.3061923&  0.59565429& -0.8955555&  1.1888545& -1.8901795&  3.07764911\\
-0.2590663&  1.36715364& -2.1170851&  2.4453460& -3.2342896&  4.11607039\\
 -1.2827794&  0.52439869& -0.9992164 & 2.1114220 &-4.0969844&  5.91743174\\
4.5782789& -0.04588379 &-1.9809570 & 0.6746645& -0.2431451 & 1.01274374\\
0.8890119 & 0.15656989& -0.6024806&  0.4886304& -0.6317594&  1.18570678\\
2.8875206& -1.88979145&  0.9503891& -0.8400636&  0.6522031 &-0.02998784
\end{pmatrix}.
\end{align*}
\end{appendices}
%%%%%%%%%%%%%%%%%%%%%%%%%%%%%%%%%

%\bmhead{Acknowledgments}
\bmhead{Acknowledgments}
This work was supported by the  Natural Sciences and Engineering Research Council of Canada  through the grant DG-2018-04449.
\bmhead{Statements and Declarations}
The authors declare they have no financial interests.
\bmhead{Authors' contributions}
C.A. worked mainly on the conceptualization and methodology, and I.S. on the simulations and applications. Both authors contributed to the  algorithm design and computer implementation. C.A. prepared the original draft and I.S. prepared the figures. Both authors reviewed and edited the manuscript. 
\bmhead{Availability of data and materials} Not applicable.
\bmhead{Code availability} The R Code used for the numerical experiments is available upon request.
\section*{Declarations}

\begin{itemize}
\item Conflict of interest: Not applicable.
\item Ethics approval : Not applicable.
\item Consent to participate: Not applicable
\item Consent for publication: Not applicable.
%\item Authors' contributions: Not applicable.
\end{itemize}




%\bibliography{stat1}% common bib file

\end{document}